\documentclass[usenatbib]{mnras}
\usepackage{amssymb} 
\usepackage{graphicx}
\usepackage{afterpage}

\title[Dust Masses in Three Supernova Remnants]{Dust masses for SN 1980K, SN1993J and Cassiopeia A from red-blue emission line asymmetries}

\author[Antonia Bevan and M. J. Barlow]{Antonia Bevan$^{1}$, M. J. 
Barlow$^{1}$ and D. Milisavljevic$^{2}$ \\ 
$^{1}$ Department of Physics and Astronomy, University 
College London, Gower Street, London WC1E 6BT, UK \\
$^{2}$ Harvard-Smithsonian Center for Astrophysics, 60 Garden Street, 
Cambridge, MA 02138, USA}

\begin{document}

\date{Accepted}

\pagerange{\pageref{firstpage}--\pageref{lastpage}} \pubyear{2016}

\maketitle

\label{firstpage}

\hyphenation{CCSNe} \hyphenation{CCSN} \hyphenation{CTIO} 
\hyphenation{VLT} \hyphenation{AAT} \hyphenation{HST}

\begin{abstract} 

We present Monte Carlo line transfer models that investigate the effects 
of dust on the very late time emission line spectra of the core 
collapse supernovae SN~1980K and SN~1993J and the 
young core collapse supernova remnant Cassiopeia~A. Their 
blue-shifted emission peaks, resulting from the removal by dust of 
redshifted photons emitted from the far sides of the remnants, and
the presence of extended red emission wings 
are used to constrain dust compositions and radii and to 
determine the masses of dust in the remnants. We estimate dust masses of 
between 0.08~--~0.15~M$_\odot$ for SN~1993J at year 16, 0.12~--~0.30~M$_\odot$ for SN~1980K at year 30 and $\sim1.1$~M$_\odot$ for Cas~A 
at year $\sim$330. Our models for the strong oxygen forbidden lines of 
Cas~A require the overall modelled profiles to be shifted to the red by 
between 700~--~1000~km~s$^{-1}$, consistent with previous estimates for the 
shift of the dynamical centroid of this remnant.

\end{abstract}


\section{Introduction}

\noindent

The large quantities of dust observed in some high redshift galaxies may 
have been produced by supernovae from massive stars, although a high dust 
formation efficiency of 0.1~--~1.0~M$_\odot$ per supernova appears to be 
required \citep{Morgan2003, Dwek2007}. Quantitative determinations of the 
masses of dust formed in the ejecta of core-collapse supernovae (CCSNe) 
have normally relied on modelling their thermal infrared dust emission 
spectra, 
\citep[][and other workers]{Dwek1983, Wooden1993, Sugerman2006, Gall2011, Gomez2012, Indebetouw2014, Matsuura2015}.
However, even with the enhanced sensitivities of modern thermal infrared 
instruments it has been difficult to detect extragalactic supernovae at 
wavelengths longwards of 8~$\mu$m more than three years after outburst.

An alternative method to determine dust masses in CCSN ejecta was proposed 
by \citet{Lucy1989} and applied by them to analyse the optical spectra of 
SN~1987A, whose H$\alpha$ and [O~{\sc i}] line profiles from two years 
after outburst onwards showed pronounced red-blue asymmetries, with their 
peak line emission shifted bluewards from line centre. This was 
interpreted and modelled as due to redshifted photons emitted from the far 
side of the ejecta suffering more absorption by newly formed dust grains 
than blue-shifted photons emitted from the near side. They noted an 
additional effect, whereby energy lost by dust-scattered photons can lead 
to a red emission wing extending to higher velocities than the largest 
velocity seen at the blue edge of the line profile.

In order to exploit the above phenomena to derive SN dust masses from 
their observed late-time emission line profiles, \citet{Bevan2016} have 
developed a Monte Carlo line transfer code, {\sc damocles}, that models 
line photons subjected to scattering and absorption by dust in expanding 
ejecta. The emitting material can have arbitrary velocity and density 
distributions and the code can handle a wide range of grain species and 
grain size distributions. \citet{Bevan2016} used {\sc damocles} to model 
the asymmetric H$\alpha$ and [O~{\sc i}] emission line profiles of 
SN~1987A between days 714 and 3604, finding that the mass of dust in the 
ejecta had increased from $\leq1.5\times10^{-3}$~M$_\odot$ on day~714 to 
$\geq0.1$~M$_\odot$ on day 3604. They also found that the presence of 
extended red emission wings in the observed line profiles at these epochs 
placed lower limits on the grain albedos, requiring grain radii 
$\geq$0.6~$\mu$m.

Blue-shifted line emission can be a common and long-lasting feature of the optical spectra of some CCSNe, with emission lines of oxygen and 
hydrogen often exhibiting red-blue asymmetries and significant 
substructure at both early times (e.g. SN~2006jc \citep{Smith2008}, SN~2005ip, SN~2006jd \citep{Stritzinger2012} and SN~2010jl \citep{Smith2012, Gall2014}) and at late times (e.g. \citealt{Milisavljevic2012}).  If these lines can be 
modelled then it may be possible to determine the masses of dust in SN 
ejecta and supernova remnants (SNRs).  This is particularly useful at late-time epochs ($\gtrsim5$ years) where
CCSNe are not currently accessible at mid-infrared and longer wavelengths. 

Of a sample of ten extragalactic CCSNe with ages of up to 50 years that 
were detected via optical spectroscopy by \citet{Milisavljevic2012}, more 
than 50\% exhibited blue-shifted emission line profiles. A range of energy 
sources can potentially illuminate such late epoch ejecta, e.g. 
irradiation by photons from a nearby OB cluster; or from the interaction 
between the forward shock and surrounding circumstellar material; 
ionization of the ejecta by a reverse shock; or photoionization by a 
central pulsar or magnetar \citep{Chevalier1992,Chevalier1994}.

In this paper we present {\sc damocles} models and dust mass estimates 
based on the late-time emission line spectra of two of the extragalactic 
supernovae observed by \citet{Milisavljevic2012} with good signal to 
noise, SN~1980K and SN~1993J. In addition, we model the asymmetric line 
profiles seen in the optical spectrum of the $\sim$330-year old 
Galactic SNR Cassiopeia~A, whose integrated spectrum was shown by 
\citet{Milisavljevic2012}, and present 
dust mass estimates for this remnant. If placed at a distance of 1~Mpc, this 
5~arcmin diameter SNR would have an angular diameter of just 
$\sim$1~arcsec, so its integrated spectrum may provide a parallel to
the spectra of extragalactic SNRs of a similar age. In Section~2 we 
describe the observational data for SN~1980K and SN~1993J and discuss the 
process of fitting their spectra. Section~3 presents models for the 
H$\alpha$ and [O~{\sc i}] profiles of SN~1980K processed by smooth 
distributions and clumped distributions of dust, while in Sections~4 and 5 
we do the same for the oxygen line profiles of SN~1993J and Cas~A, 
respectively. We summarise our conclusions in Section~6.

\begin{figure*} 
\centering 
\includegraphics[clip=true,scale=0.6, trim=20 0 50 20]{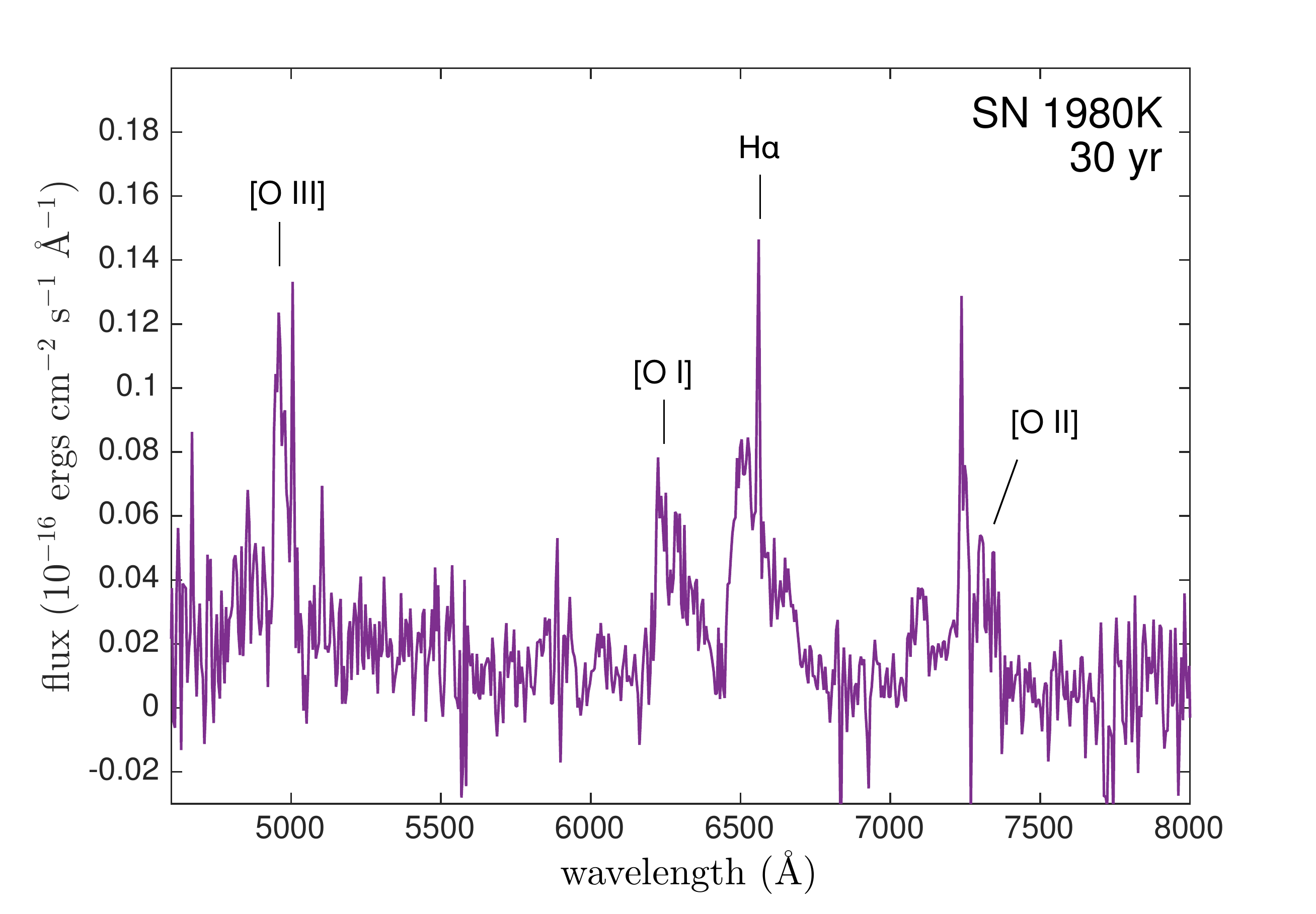}
\includegraphics[clip=true,scale=0.6, trim=30 0 50 20]{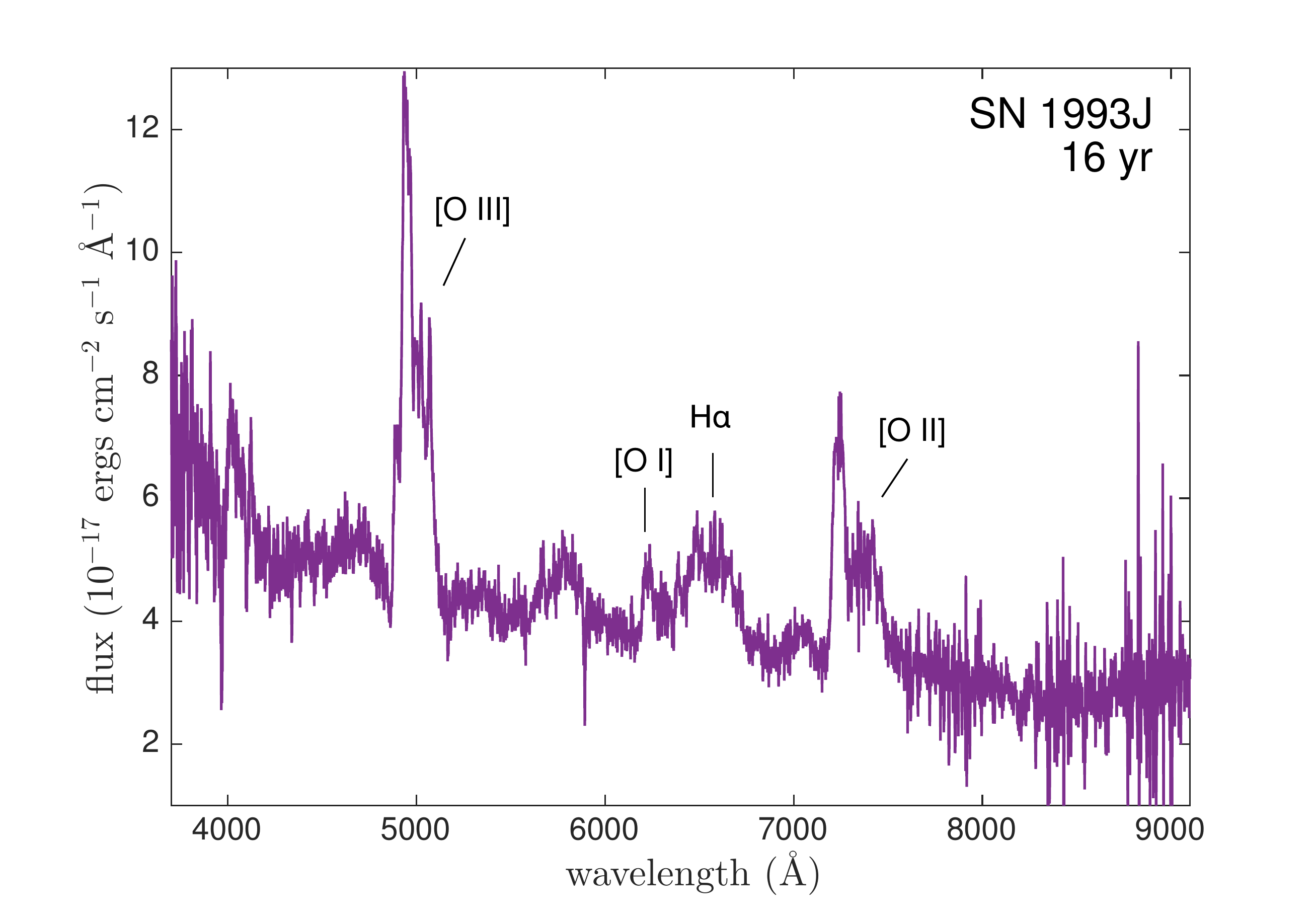}
\caption{{\em Above:} The optical spectrum of SN~1980K on 9 October 2010 
at 30 years post-explosion.  {\em Below:} The optical spectrum of SN~1993J 
on 9 December 2009 at 16 years post-explosion. Both spectra were obtained 
by \citet{Milisavljevic2012}.} 
\label{spectra} 
\end{figure*}

\section{SN~1980K and SN~1993J}

SN~1980K is located in NGC 6946, at a distance of 
approximately 5.9~Mpc \citep{Karachentsev2000}.  It was discovered by 
P. Wild on 28 October 1980 and had reached a peak brightness of $V=11.4$ 
mag by November that year \citep{Buta1982}.  The detection of a broad 
H$\alpha$ line in early spectra and a linearly decaying light curve after 
peak brightness resulted in its classification as a Type IIL supernovae 
\citep{Barbon1982}.  SN~1980K continued to decline steadily in the optical 
although it was still detected almost seven years after maximum light by 
narrow passband imaging by \citet{Fesen1988}, whose follow-up low dispersion 
observations found that the spectra exhibited broad H$\alpha$ and 
[O~{\sc i}] $\lambda\lambda$6300,6363 emission, with other weaker 
optical lines also present.

Spectroscopic and photometric observations of SN~1980K have revealed a 
very slow monotonic fading over a period of $\sim$20 years.  This 
unusually slow rate of decline suggested that the observations may in fact 
be a product of light echoes scattering off and heating circumstellar 
material.  This was first suggested 
by \citet{Chevalier1986} based on observations during the 
first year after outburst.  From further modelling and analyses of 
late-time observations \citet{Sugerman2012} concluded that light echoes 
were indeed present and that the evolution of the observations could be 
explained by scattered and thermal echoes off a thin circumstellar shell 
of dust of mass $\lesssim 0.02$~M$_{\odot}$ approximately 14~--~15 
lightyears from the progenitor. Of particular relevance was their 
discussion of the origin of the broad, high velocity H$\alpha$ and [O~{\sc i}]
$\lambda\lambda$6300,6363 lines which were not present in 
early spectra taken during the first two years.  They concluded that the 
shape of these lines could not be a product of a light echo since the high 
velocities seen in the late-time spectra were not present in early 
spectra.

In 1981, the emergence of a near-IR flux excess had provided the first 
indications of dust in the ejecta of SN~1980K \citep{Dwek1983}.  However, 
it could not be confirmed whether this excess infrared (IR) flux was the 
result of newly-formed dust condensing in the ejecta or came from pre-existing 
grains located in a circumstellar shell illuminated by radiation from the 
outburst.  In addition to the detection of the signature emission from hot 
dust grains in the near-IR, highly blue-shifted line profiles in the 
optical spectra of SN~1980K have been observed for a number of years 
after \citep{Fesen1990,Fesen1994,Fesen1995,Fesen1999}.  The presence of 
dust in its ejecta was postulated by \citet{Milisavljevic2012} based on 
the observed blue-shifting of the optical line profiles, still 
present even in very late-time spectra (30 years). It is these late 
blue-shifted line profiles of SN~1980K at 30 years that we have modelled 
and present here.  An explosion date of 2 October 1980 \citep{Montes1998} 
was adopted for all models.

SN~1993J is a very well-observed supernova and is only 
surpassed by SN~1987A in regards to the quality and frequency of its 
observations.  It is located in the nearby M~81 galaxy, 3.6~Mpc away 
\citep{Freedman1994} and was discovered on 28 March 1993 
\citep{Ripero1993}.  It reached a maximum brightness of $V=10.8$ mag 
making it the brightest supernova in the northern hemisphere since 
SN~1954A. Early spectra showed typical Type II features including 
broad H$\alpha$ emission. However, its 
evolution was atypical and the appearance of He lines in later spectra 
resulted in its classification as Type IIb.  The similarities to Type Ib 
and Type Ic supernovae were noted however and this supernova has been
important for understanding the relationship between the Type I and Type 
II CCSN categories \citep{Fillipenko1993,Garnavich1993}.  Extensive 
reviews of SN~1993J are given by \citet{Wheeler1996}, who cover the early 
evolution of the object, and by \citet{Matheson2000a,Matheson2000b} who 
discuss the later evolution of the optical spectra.

The relatively isolated and nearby position of SN~1993J led to regular 
monitoring at X-ray, radio and optical wavelengths. We were particularly 
interested in late-time optical spectra obtained at 16 years post-outburst 
and the presence, or otherwise, of dust in the ejecta as postulated by 
\citet{Fransson2005} and \citet{Milisavljevic2012}.  An explosion date of 
27 March 1993 \citep{Baron1993} is adopted for all models.

\subsection{Fitting the Late-Time Optical Spectra of SN~1980K and SN~1993J}
\label{80K_93J_models}

Late-time spectra of both SN~1980K and SN~1993J were published by 
\citet{Milisavljevic2012} and we present them here in Figure 
\ref{spectra}.

The spectra of SN 1980K were obtained on 9 October 2010 using the 2.4m 
Hiltner telescope at the MDM observatory. The Mark III Spectrograph was used with a 
SITe $1024 \times 1024$ CCD detector and a $1.2'' \times 4.5'$ slit. 
Exposures were $2 \times 3000$~s and spectra spanned the wavelength range 
4600~--~8000~\AA\ with a spectral resolution of 7~\AA. The spectrum 
presented in Figure \ref{spectra} is of SN~1980K at approximately 30 years 
after outburst. Significant blue-shifting can be seen in virtually all 
lines, but especially in the H$\alpha$ and [O~{\sc i}] 
$\lambda$6300,6363 lines which exhibit a pronounced flux bias 
towards the blue and a strongly blue-shifted peak (see Figure 
\ref{spectra}).  Narrow nebular lines of H$\alpha$ and [O~{\sc iii}] 
provide useful rest velocity reference points and the small recession 
velocity of 40~km~s$^{-1}$ of NGC~6946 has been corrected for.

The optical spectrum of SN 1993J was obtained on 9 December 2009 with the 
6.5m MMT at Mt. Hopkins in Arizona using the 
HECTOSPEC optical fibre fed spectrograph. Spectra from the $1.5''$ 
diameter fibres covered the wavelength range of 3700~--~9200~\AA\ with a 
full-width at half maximum (FWHM) resolution of 5~\AA.  The total exposure 
time was 3600~s. The observations were obtained as a part of a survey of 
the supernova remnants in M81.  SN~1993J was approximately 16 years old 
when the spectra were obtained.  Many of the lines in its optical spectrum 
exhibit a flux bias towards the blue and also display noticeable 
substructure (see Figure \ref{spectra}).  They are generally broad and 
there is a significant degree of blending between lines.  The lines least 
blended with other lines are those of [O~{\sc iii}] 
$\lambda\lambda$4959,5007 and [O~{\sc ii}] 
$\lambda\lambda$7319,7330 which both demonstrate significantly 
asymmetrical profiles, with both the flux and the peak of their profiles 
shifted towards the blue. The spectra have been corrected for an M~81 
velocity of $-140$~km~s$^{-1}$ \citep{Matheson2000b}.

\begin{figure} 
\centering 
\includegraphics[clip=true,scale=0.42,trim= 20 0 50 0]{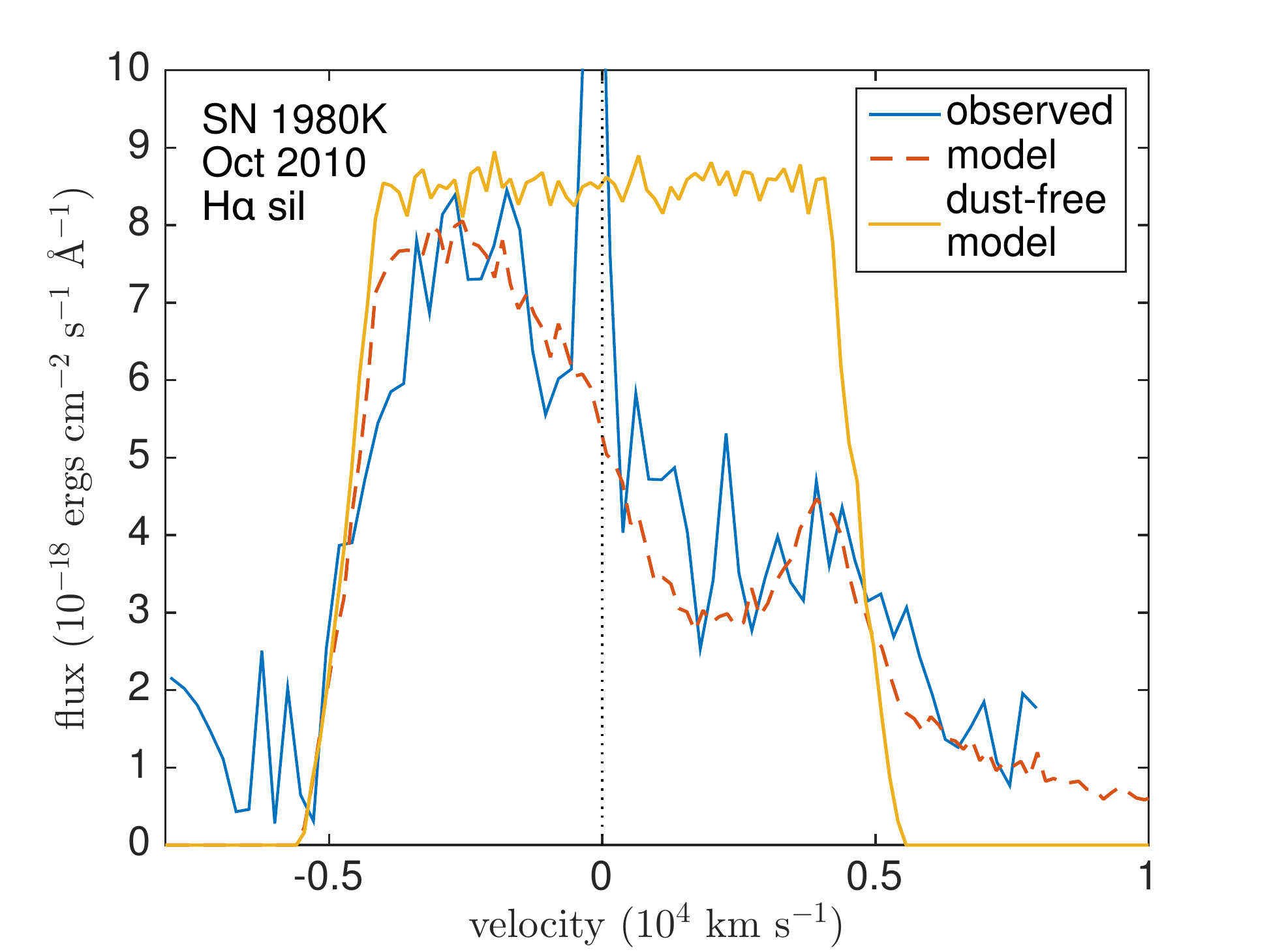} 
\includegraphics[clip=true,scale=0.42,trim= 17 0 50 0]{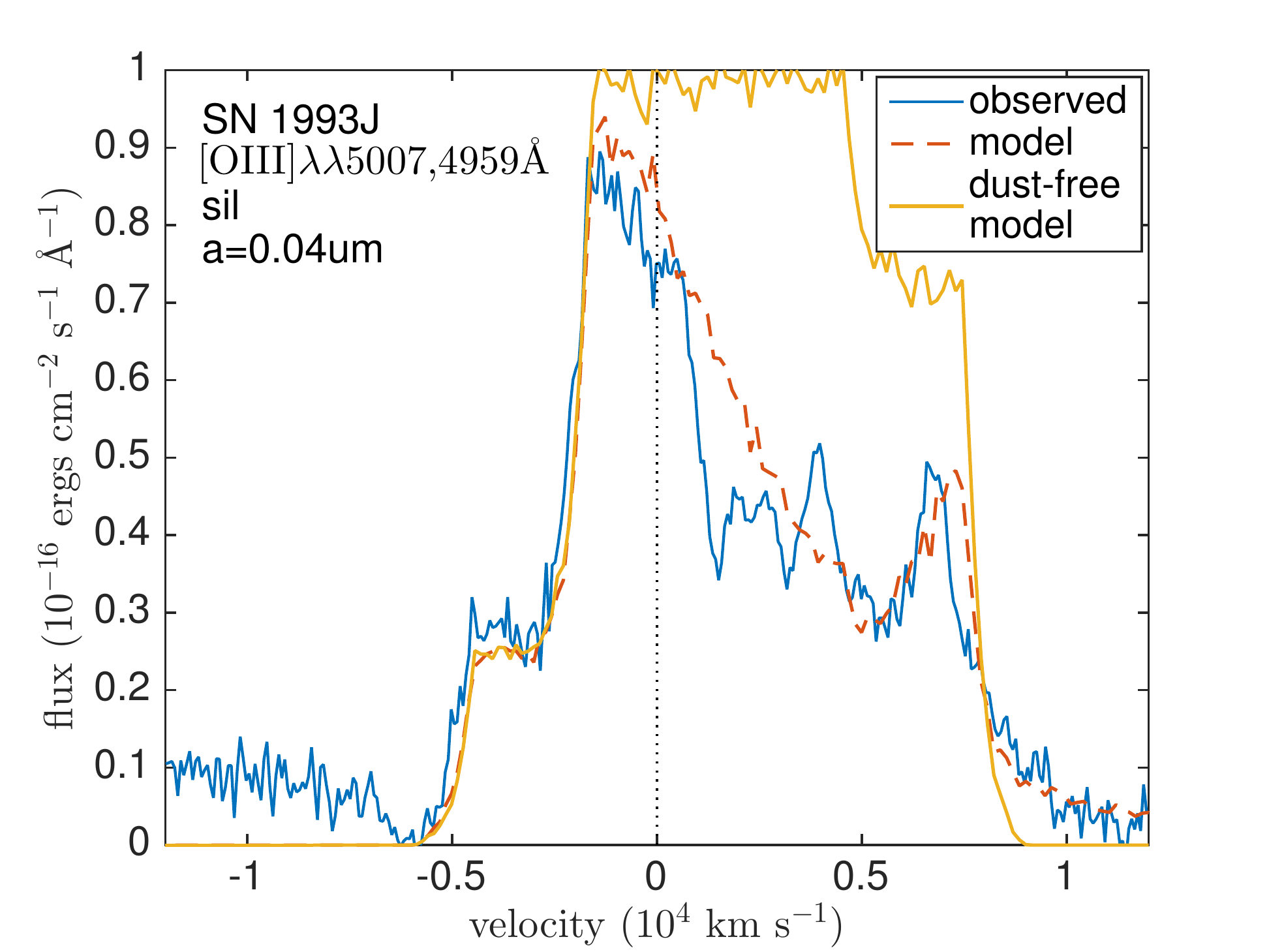} 
\caption{Best-fitting smooth dust 
models along with the intrinsic dust-free boxy profile for the H$\alpha$ 
line profile for SN~1980K {\em (upper)} and the [O~{\sc iii}] 
$\lambda\lambda$4959,5007 line profile for SN~1993J {\em 
(lower)}.  The intrinsic dust-free modelled line profile is given in 
yellow, the dust-affected modelled line profile in red and the observed 
line profile in blue.} 
\label{boxy} 
\end{figure}


Our modelling of SN~1980K focussed on the H$\alpha$ line and the [O~{\sc i}]
$\lambda\lambda$6300,6363~\AA\ doublet.  Both of these line profiles 
exhibited a very strong blue-shifted asymmetry indicative of the presence 
of dust in the ejecta and, like SN~1987A, were sufficiently distinct that 
they provided the best options for modelling purposes.  Other lines were 
either too blended with each other or too noisy to be reliable.

SN~1993J exhibited its strongest line asymmetries in the oxygen lines, and 
in particular we focussed our modelling on the [O~{\sc ii}]
$\lambda\lambda$ 7319,7330 and [O~{\sc iii}]
$\lambda\lambda$4959,5007 doublets mentioned above.  The [O~{\sc i}]
$\lambda\lambda$ 6300,6363 doublet was not modelled for SN~1993J as 
it was quite blended with the H$\alpha$ line and the two features could 
not be easily separated.  The [O~{\sc ii}] and [O~{\sc iii}] lines, being 
more distinct, were therefore the more sensible candidates for modelling.

\afterpage{ 
\centering 
\setlength{\tabcolsep}{4pt} 
\begin{table*} 
\centering 
  \caption{The parameters used for the smooth and clumped models of 
SN~1980K for media composed of 100\% amorphous carbon dust grains of 
radius $3.5$~$\mu$m, or 100\% silicate dust grains of radius $0.1$~$\mu$m.  
Optical depths are given from $R_{in}$ to $R_{out}$ at $\lambda = 
6300$~\AA\ for [O~{\sc i}] and $\lambda = 6563$~\AA\ for H$\alpha$.  The 
doublet ratio for [O~{\sc i}] $\lambda\lambda$6300,6363 was fixed 
to be 3.1.  Smooth dust models are listed in the first four rows and 
clumped dust models in the last four rows.}
	\label{80K}
	\centering
  	\begin{tabular}{@{} ccccccccccccccc @{}}
    	\hline
  & clumped? & species &$a$&$V_{max}$ & $V_{min}$ & $R_{in}/R_{out}$ & 
$\beta$ & $R_{out}$ & $R_{in}$ & doublet & $\tau_{\lambda}$ & $f$ & 
$R_{clump}$ &$M_{dust}$ \\
	&& &($\mu$m)&(km~s$^{-1}) $& (km~s$^{-1} $)& && (10$^{17}$cm) & 
(10$^{17}$cm) &ratio&&&(10$^{17}$cm) &($M_{\odot}$)  \\
	\hline
	H$\alpha$ &no& sil & 0.1&5500 & 4125 & 0.75 & 2.0 & 5.2 & 3.9 & - & 1.41 & - & - &0.10\\ \relax 
H$\alpha$ &no&amC& 3.5&5500 & 4125 & 0.75 & 2.0 & 5.2 & 3.9 & - & 0.57 & - & -& 0.30\\ \relax 
[O~{\sc i}] &no&sil&0.1& 5500 & 4125 & 0.75 & 4.0 & 5.2 & 3.9 & 3.1 & 2.81 & - & - & 0.20\\ \relax 
[O~{\sc i}] &no&amC&3.5& 5500 & 4125 & 0.75 & 4.0 & 5.2 & 3.9 & 3.1 & 1.24 & - & -& 0.65\\ \\ 
H$\alpha$ &yes&sil & 0.1&5500 & 4125 & 0.75 & 2.0 & 5.2 & 3.9 & - & 1.68 & 0.1 & 0.2 & 0.12\\ \relax 
H$\alpha$ &yes&amC& 3.5&5500 & 4125 & 0.75 & 2.0 & 5.2 & 3.9 & - & 0.73 & 0.1 & 0.2& 0.38\\ \relax 
[O~{\sc i}] &yes&sil&0.1& 5500 & 4125 & 0.75 & 4.0 & 5.2 & 3.9 & 3.1 & 2.81 & 0.1 & 0.2 & 0.30\\ \relax 
[O~{\sc i}] &yes&amC&3.5& 5500 & 4125 & 0.75 & 4.0 & 5.2 & 3.9 & 3.1 & 1.72 & 0.1 & 0.2& 0.90\\
    \hline
\end{tabular} 
\end{table*} 

\begin{figure} 
\centering 
\includegraphics[scale=0.38,clip=true, trim=20 0 40 20]{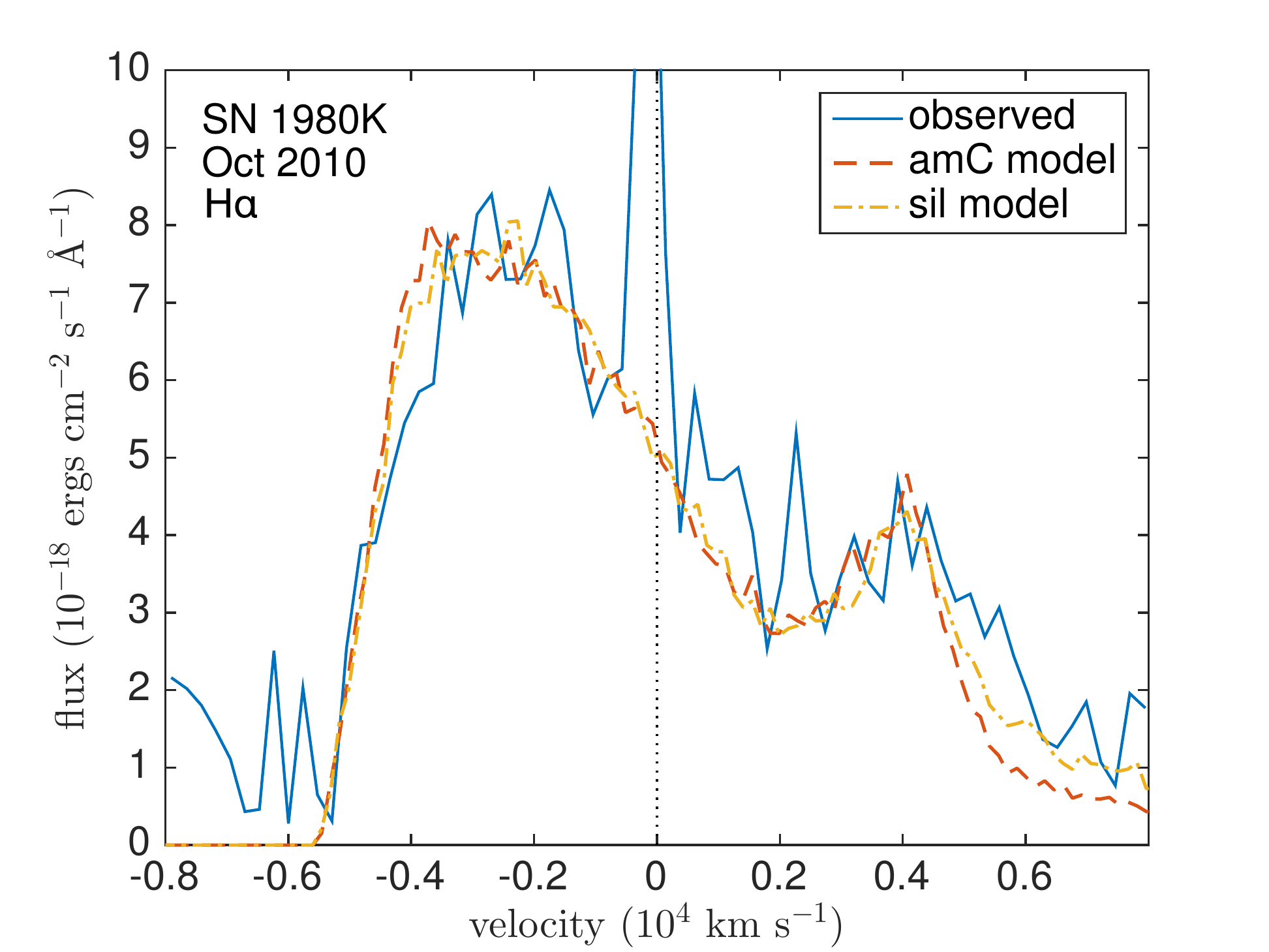}

\includegraphics[scale=0.38,clip=true, trim=20 0 40 20]{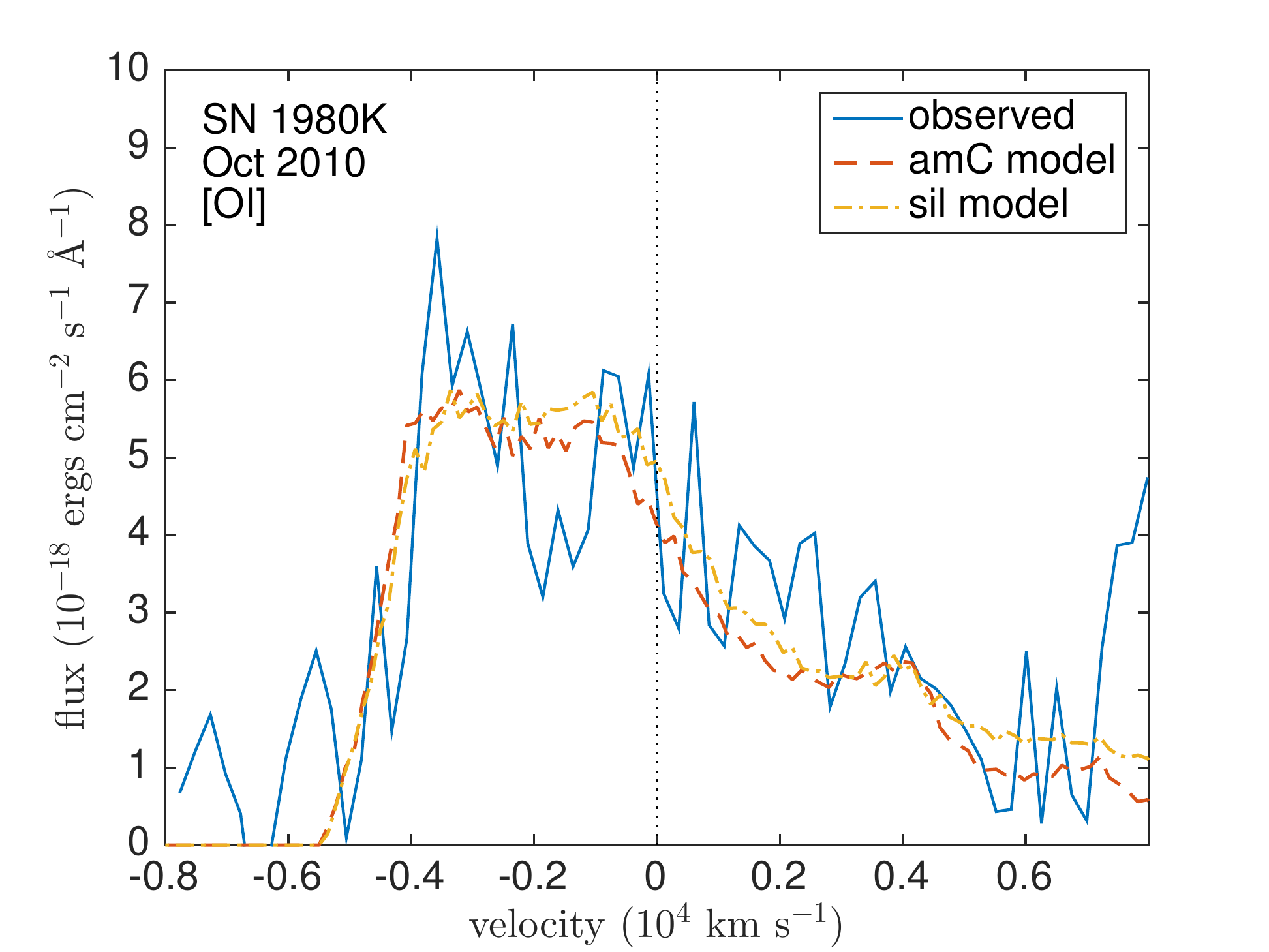} 
\caption{Best smooth dust fits for the SN~1980K H$\alpha$ line ({\em upper})  
and for the [O~{\sc i}] $\lambda\lambda$6300,6363 doublet ({\em lower}) 
for the parameters detailed in Table \ref{80K}.  Smooth dust fits with 
astronomical silicate grains of radius $a=0.1$~$\mu$m are presented in yellow and smooth dust fits with amorphous carbon grains of radius 
$a=3.5$~$\mu$m in red.  For the [O~{\sc i}] doublet, 
zero velocity was set at $\lambda=6300$~\AA.}
\label{80K_smooth} 
\end{figure}
}

Our approach to modelling the line profiles of both SN~1980K and SN~1993J 
followed the same principles as described for SN~1987A by 
\citet{Bevan2016}. The maximum velocity was identified from 
the point at which flux vanishes on the blue side, the inner to outer 
radius ratio determined from various inflection points and the density 
profile was determined from the shape of the profile. 
We began the modelling by considering a smooth, coupled 
distribution of dust and gas before moving on to consider the effect on 
these models of including a clumped dust geometry whilst maintaining a 
smooth gas emissivity distribution. We first examined the line profiles 
in order to determine the maximum and minimum velocities before moving 
on to establish approximately the exponent of the dust and gas density 
distributions. Having fixed the starting values for these quantities, we 
iterated over the grain size and dust mass in order to fit the profile.  
We also occasionally varied the other parameters in order to optimise the 
fits to the data. We assumed that the oxygen doublets were optically thin 
for both SN~1980K and SN~1993J and using the theoretical transition 
probabilities detailed by \citet{Zeippen1987} and \citet{Storey2000}
adopted a constant intrinsic flux ratio of 3.1 between the [O~{\sc i}] 
$\lambda\lambda$6300,6363 components, 1.2 between the [O~{\sc ii}] 
$\lambda\lambda$7319,7330 components and 2.98 between the [O~{\sc iii}]
$\lambda\lambda$5007,4959 components.
The intrinsic line profiles for SN~1980K and SN~1993J  
prior to dust effects both needed to be very `boxy', that is, the ratio of 
the inner to outer radii is very high so that the overall profile has a 
very square shape (see Figure \ref{boxy}). We present the intrinsic 
dust-free profile along with the best-fitting smooth H$\alpha$ model for 
SN~1980K in the top pane of Figure \ref{boxy}, along with the intrinsic 
dust-free profile, and the best-fitting smooth [O~{\sc iii}]
$\lambda\lambda$4959,5007 model for SN~1993J in the lower pane
of Figure \ref{boxy}.

\begin{figure*} 
\centering 
\includegraphics[scale=0.38,clip=true, trim=20 0 40 20]{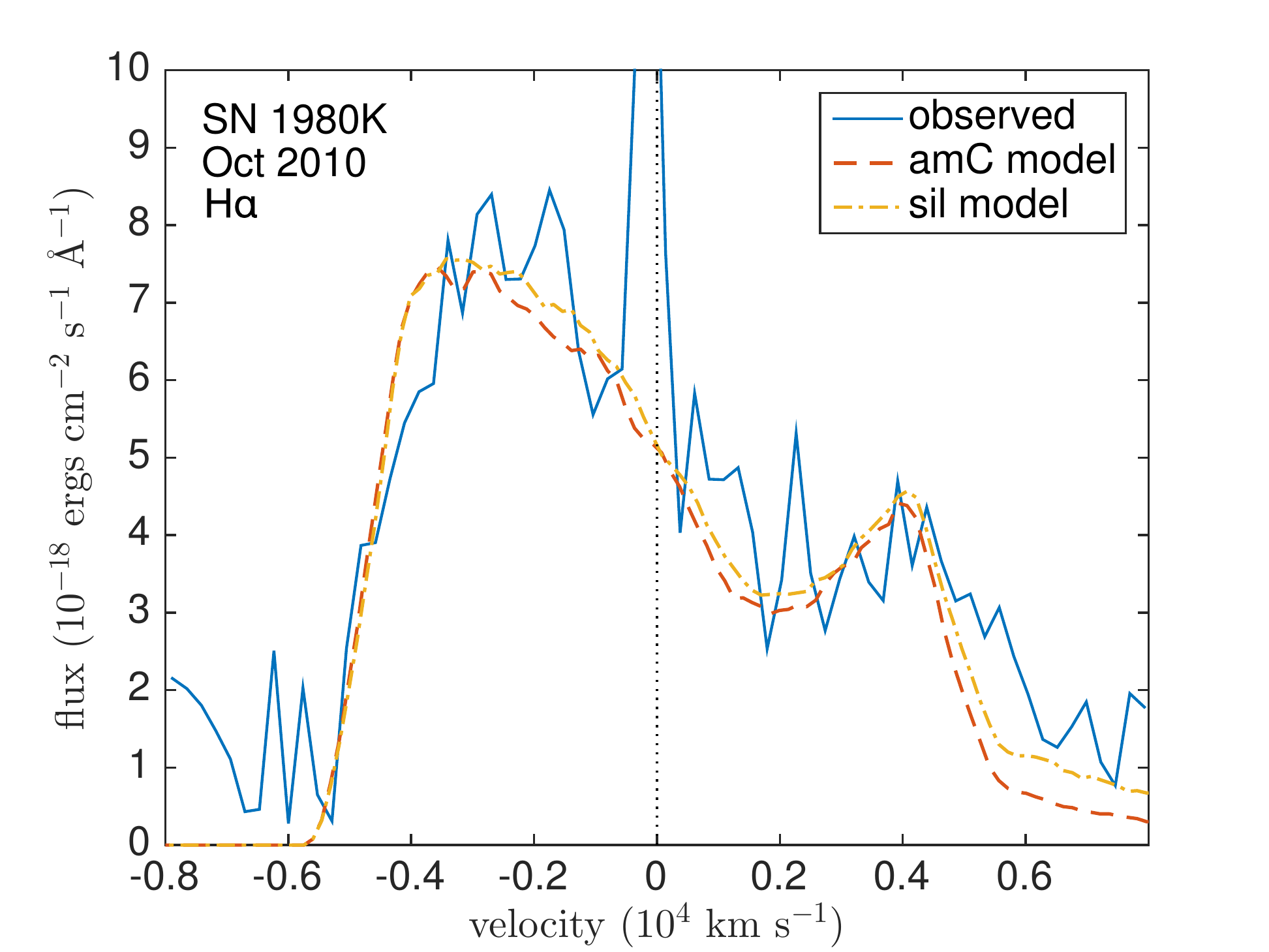}
\includegraphics[scale=0.38,clip=true, trim=20 0 40 20]{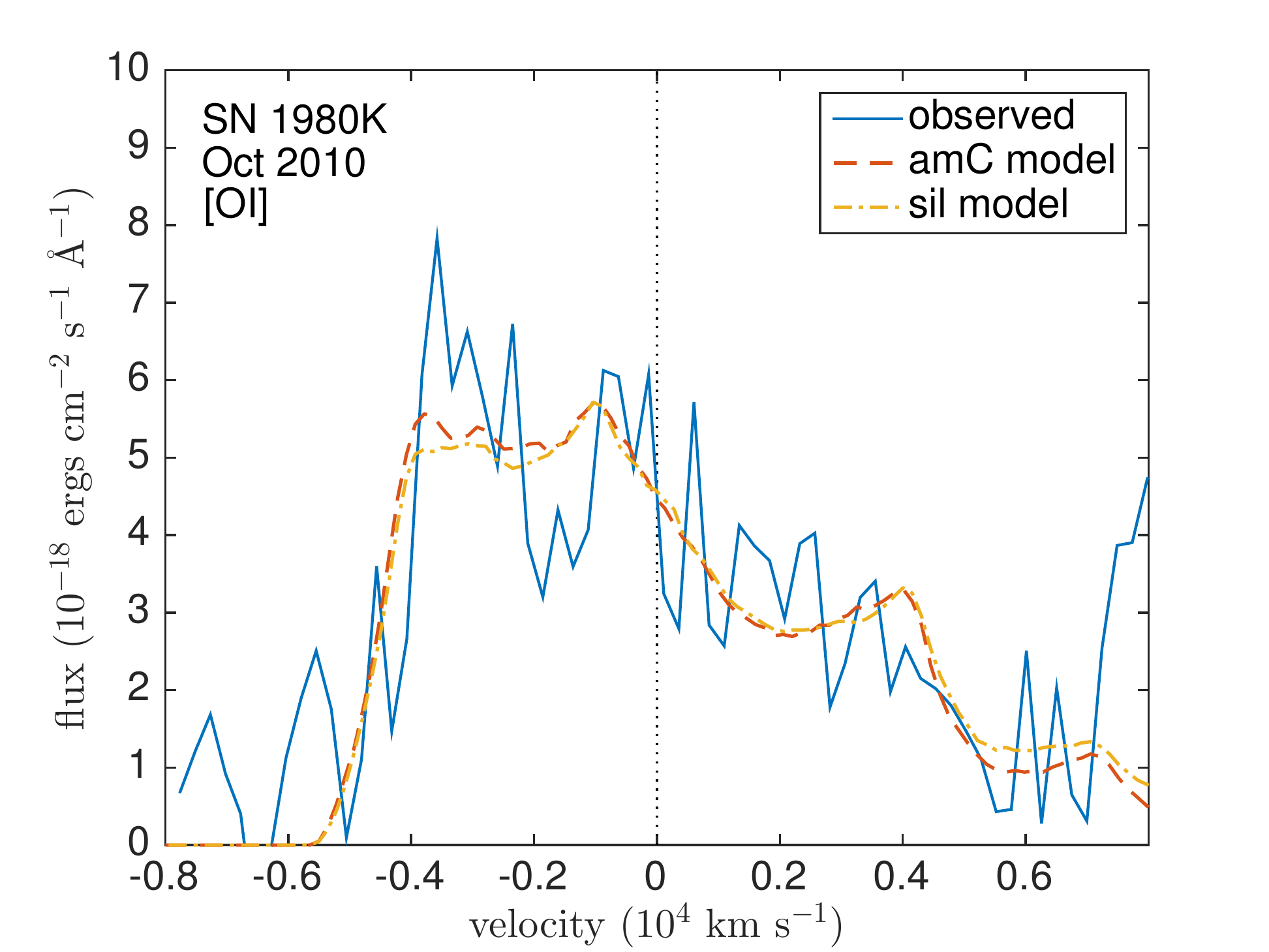} 
\caption{Best clumped dust fits to the 
SN~1980K H$\alpha$ line ({\em top}) and the [O~{\sc i}]
$\lambda\lambda$6300,6363 doublet ({\em bottom}) for the 
parameters detailed in Table \ref{80K}.  Clumped dust fits with 
astronomical silicate grains of radius $a=0.1$~$\mu$m are presented on the 
left and clumped dust fits with amorphous carbon grains of radius 
$a=3.5$~$\mu$m are presented on the right. For the [O~{\sc i}] doublet, 
zero velocity was set at $\lambda=6300$~\AA.} 
\label{80K_clumped} 
\end{figure*}

The parameters for the smooth and clumped dust fits that we obtained for 
SN~1980K and SN~1993J are detailed in Tables \ref{80K} and \ref{93J} 
respectively. The smooth dust line profile fits for SN~1980K are presented 
in Figure \ref{80K_smooth} and the clumped dust line profile fits are 
presented in Figure \ref{80K_clumped}.  The smooth dust line profile fits 
for SN~1993J are presented in Figure \ref{93J_smooth} and the clumped dust 
line profile fits are presented in Figure \ref{93J_clumped}.

\section{Models for the Year 30 line profiles of SN~1980K}
\subsection{SN~1980K Smooth Dust Models}
\label{80Ksmooth}

We obtained good fits to both the H$\alpha$ line and the [O~{\sc i}] 
$\lambda\lambda$6300,6363 doublet from SN~1980K (see Figure 
\ref{80K_smooth}).  In particular, an extended wing on the red side of the 
profile was seen in both cases. This was more important for the H$\alpha$ 
line since we could be sure that it was not a product of blending with an 
adjacent broad line (the presence of an extended red wing in the [O~{\sc 
i}] doublet may be due to blending with the blue wing of the H$\alpha$ 
line).  The H$\alpha$ red wing allowed us to place constraints on the 
dust albedo using the approach described by \citet{Bevan2016}.
A high albedo of $\omega\approx0.8$ was required to reproduce the 
flux in the region between $+6000$ and $+8000$~km~s$^{-1}$.  Astronomical silicate grains  of radius $a=0.1$~$\mu$m have an albedo
of this magnitude at this wavelength   \citep{Draine1984}, but amorphous carbon grain are never 
this reflective regardless of their size.  As well as the best-fitting 
silicate model in Figure \ref{80K_smooth}, we include line profile fits 
using a large grain size ($a=3.5$~$\mu$m) for amorphous carbon to generate 
as high an albedo as possible ($\omega\approx0.6$) illustrating the 
slightly worse fit to the H$\alpha$ line.

\subsection{SN~1980K Clumped Dust Models}

Motivated by our modelling of SN~1987A \citep{Bevan2016}, we adopted a clumped 
dust structure with a clump volume filling factor of $f=0.1$ and clumps of 
radius $R_{clump}=R_{out}/25$. All the dust was located in clumps but the 
gas emission remained distributed smoothly according to the distribution 
derived for the smooth models.  A summary of the parameters for the 
best-fitting clumped models is presented in Table \ref{80K} and the fits 
are presented in Figure \ref{80K_clumped}.

\subsection{SN~1980K Discussion} 
\label{scn:discuss}
 
 The models for the H$\alpha$ and [O~{\sc i}] $\lambda\lambda$6300,6363 
profiles are broadly consistent with each other. The primary differences 
in their derived parameters (Table \ref{80K}) are in the exponents of the 
density distributions and the total dust masses, with the oxygen 
distribution following a steeper density trend than the more diffusely 
emitted hydrogen.  In a similar manner to the early phase models of 
SN~1987A, the [O~{\sc i}] $\lambda\lambda$6300,6363 line profile models 
required significantly greater dust masses than the H$\alpha$ models. We 
believe that this is likely to be due to the same reason discussed for 
SN~1987A by \citet{Bevan2016}, namely that the dust forming regions may be 
more concentrated towards those zones which are oxygen-rich 
\citep{Kozma1998a}.  As a result, it seems possible that if most of the 
gas-phase oxygen is located in clumps along with the dust then the 
discrepancy in the dust masses could potentially be resolved by 
considering more complex, decoupled distributions of dust and gas with 
diffuse hydrogen emission and clumped oxygen emission. \citet{Bevan2016} 
illustrated this possibility for SN~1987A. We note that for the clumped 
models for both SN~1980K and SN~1993J the difference between the dust 
masses derived from the [O~{\sc i}] $\lambda\lambda$6300,6363 fits and 
H$\alpha$ fits is around a factor of approximately two, very similar to 
that seen for SN~1987A.

As discussed in Section \ref{80Ksmooth} above,
the H$\alpha$ and [O~{\sc i}] line profiles of SN~1980K both exhibit an 
extended scattering wing which requires dust with a 
high albedo to fit it.  Amorphous carbon models do not fit the red side of 
the H$\alpha$ profile very well, even for very large grain sizes. We 
therefore adopt a silicate dust composition, which fits the profiles 
somewhat better. We note that a combination of grain species would also be 
capable of producing the high albedo that is required but would be 
expected to result in dust masses somewhere in between those of the 
amorphous carbon and silicate models.  The relatively low signal-to-noise 
ratio of both profiles means that a small degree of variation in the 
parameters is found to generate modelled line profiles that also fit the 
data reasonably.  This is important for the determination of the high 
albedo which is based on a relatively small section of the observed line 
profile in the red wing of the data.  Further observations with a higher 
signal-to-noise ratio could be beneficial.

\afterpage{ 
\centering 
\setlength{\tabcolsep}{4pt} 
\begin{table*} 
\centering 
	\caption{The parameters used for the smooth and clumped 
Year16 models for SN~1993J for media composed of 100\% amorphous carbon 
dust grains of 
radius $3.5$~$\mu$m, or 100\% silicate dust grains of radius $0.1$~$\mu$m.  
Optical depths are given from $R_{in}$ to $R_{out}$ at $\lambda = 
7319$~\AA\ for [O~{\sc ii}] and $\lambda = 4959$~\AA\ for [O~{\sc iii}].  
Smooth dust models are listed in the first four rows and clumped dust 
models in the last four rows.}
	\label{93J}
	\centering
  	\begin{tabular}{@{} ccccccccccccccc @{}}
    	\hline
  & clumped? & species &$a$&$V_{max}$ & $V_{min}$ & $R_{in}/R_{out}$ & 
$\beta$ & $R_{out}$ & $R_{in}$ & doublet & $\tau_{\lambda}$ & $f$ & 
$R_{clump}$& $M_{dust}$ \\
	&& &($\mu$m)&(km~s$^{-1}) $& (km~s$^{-1} $)& & & (10$^{17}$cm) & 
(10$^{17}$cm) &ratio&&&(10$^{17}$cm) & ($M_{\odot}$) \\
	\hline
	{[O~{\sc iii}]} &no&sil & 0.04&6000 & 4500 & 0.75 & 7 & 3.2 & 2.4 & 2.98 & 0.65 & - & - & 0.10\\ \relax 
{[O~{\sc iii}]} &no&amC&0.2& 6000 & 4500 & 0.75 & 7 & 3.2 & 2.4 & 2.98& 0.63 & - & -& 0.005\\ \relax 
{[O~{\sc ii}]} &no&sil& 0.1&6000 & 4500 & 0.75 & 7 & 3.2 & 2.4 & 1.23 & 0.74 & - & - & 0.05\\ \relax 
{{[O~{\sc ii}]}} &no&amC&3.5& 6000 & 4500 & 0.75 & 7 & 3.2 & 2.4 & 1.23 & 0.60 & - & - & 0.12\\ \\ 
{[O~{\sc iii}]} &yes&sil &0.04& 6000 & 4500 & 0.75 & 7 & 3.2 & 2.4 & 2.98 &1.00 & 0.1 & 0.13& 0.15 \\ \relax 
{[O~{\sc iii}]} &yes&amC&0.2& 6000 & 4500 & 0.75 & 7 & 3.2 & 2.4 & 2.98 &0.96 & 0.1 & 0.13& 0.008\\ \relax 
{[O~{\sc ii}]} &yes&sil&0.1 &6000 & 4500 & 0.75 & 7 & 3.2 & 2.4 & 1.23 & 1.12 & 0.1 & 0.13 &0.08\\ \relax 
{[O~{\sc ii}]} &yes&amC& 3.5& 6000 & 4500 & 0.75 & 7& 3.2 & 2.4 & 1.23 & 0.95 & 0.1 & 0.13 & 0.18 \\
    \hline
  \end{tabular}

\end{table*} 
}

There has been only limited discussion of the dust mass that could be 
present in the ejecta of SN~1980K other than that dust must be present 
based on the asymmetrical optical line profiles. The initial evidence for 
dust in the ejecta of SN~1980K was based on the near-IR flux excess seen a 
few hundred days after outburst \citep{Dwek1983}.  They discussed the 
possibility of newly-formed dust in the ejecta accounting for this IR flux 
but also acknowledged the possibility that the excess IR flux could also 
be a product of the reheating of pre-existing dust grains in circumstellar 
material. \citet{Sugerman2012} presented detailed modelling of the light 
echoes of SN~1980K and concluded that thermal echoes off a thin 
circumstellar shell of dust, of light emitted from an ultraviolet flash in 
the first two days, or optical light emitted in the first 150 days, were 
contributing to the observed IR flux but could only account for a small 
fraction of that flux.  Whilst circumstellar dust may in fact explain the 
early IR emission, dust located in a circumstellar shell outside the 
supernova ejecta cannot explain the observed asymmetries in the late-time 
optical line profiles.

Other explanations for the stubborn presence of strongly blue-shifted 
asymmetrical optical lines have been advanced previously for SN~1980K. 
\citet{Fesen1990} argued that broad asymmetrical lines in the early 
spectra could arise as a result of the impact between the blast wave and 
pre-existing circumstellar material.  Similarly, a `clumpy wind' model with emission coming from shocked clumps in 
order to explain the blue-shifted lines was put forward by 
\citet{Chugai1994}.  Both of these mechanisms could 
theoretically result in asymmetrical line profiles as a result of the 
emission from the approaching side of the supernova ejecta reaching us 
before emission from the receding side.  However, both of these 
suggestions were ruled out by \citet{Sugerman2012} based primarily on 
analyses of the various time scales involved but also on their inability 
to reproduce the observed late-time excess IR flux.  \citet{Fesen1990} 
also noted the possibility of blue-shifted lines arising as a result of 
dust forming in the ejecta but were doubtful as to the feasibility of the 
diffusely emitted hydrogen being so strongly affected by dust forming in 
the more dense, central regions of the ejecta.  \citet{Sugerman2012} 
estimated that a dust mass of $\sim10^{-3}$~M$_{\odot}$ was needed to 
explain the mid-IR spectral energy distribution at similar epochs 
(23~--~30 years post-outburst) to 
those investigated here, with the presence of as much as a few M$_{\odot}$ 
of cold dust possible due to the fact that the SED was still rising at 
24~$\mu$m.  This latter possibility was noted based on the depth to which 
{\em Herschel} could probe during far-IR observations of NGC~6946 in 2010. 
A few M$_{\odot}$ of cold dust in the ejecta of SN~1980K would not have 
been detected by {\em Herschel}.  These results, though not particularly 
constraining, are consistent with our current estimates that up to 
0.9~M$_{\odot}$ of dust can be present in the ejecta of SN~1980K. Given 
the strong oxygen forbidden lines in the spectrum of SN~1980K, a 
silicate-dominated dust composition, rather than amorphous carbon, seems 
likely, implying a clumped dust mass of between 0.12 and 0.30~M$_{\odot}$ 
at year 30 (Table \ref{80K}).

\section{Models for the Year 16 line profiles of SN~1993J}
\subsection{SN~1993J Smooth Dust Models}

\begin{figure}
\centering 
\includegraphics[scale=0.38,clip=true, trim=10 0 40 20]{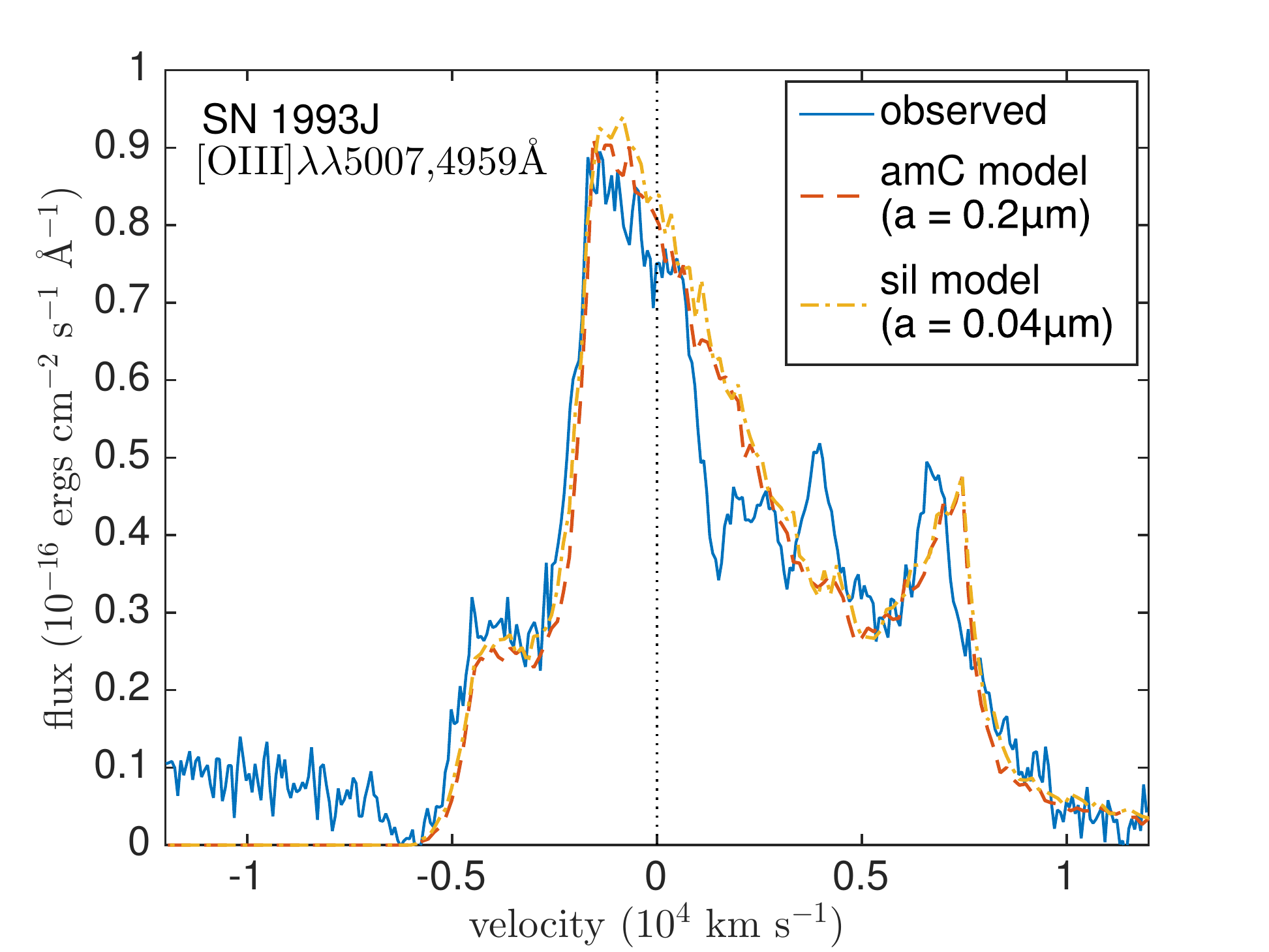}
\includegraphics[scale=0.38,clip=true, trim=0 0 40 20]{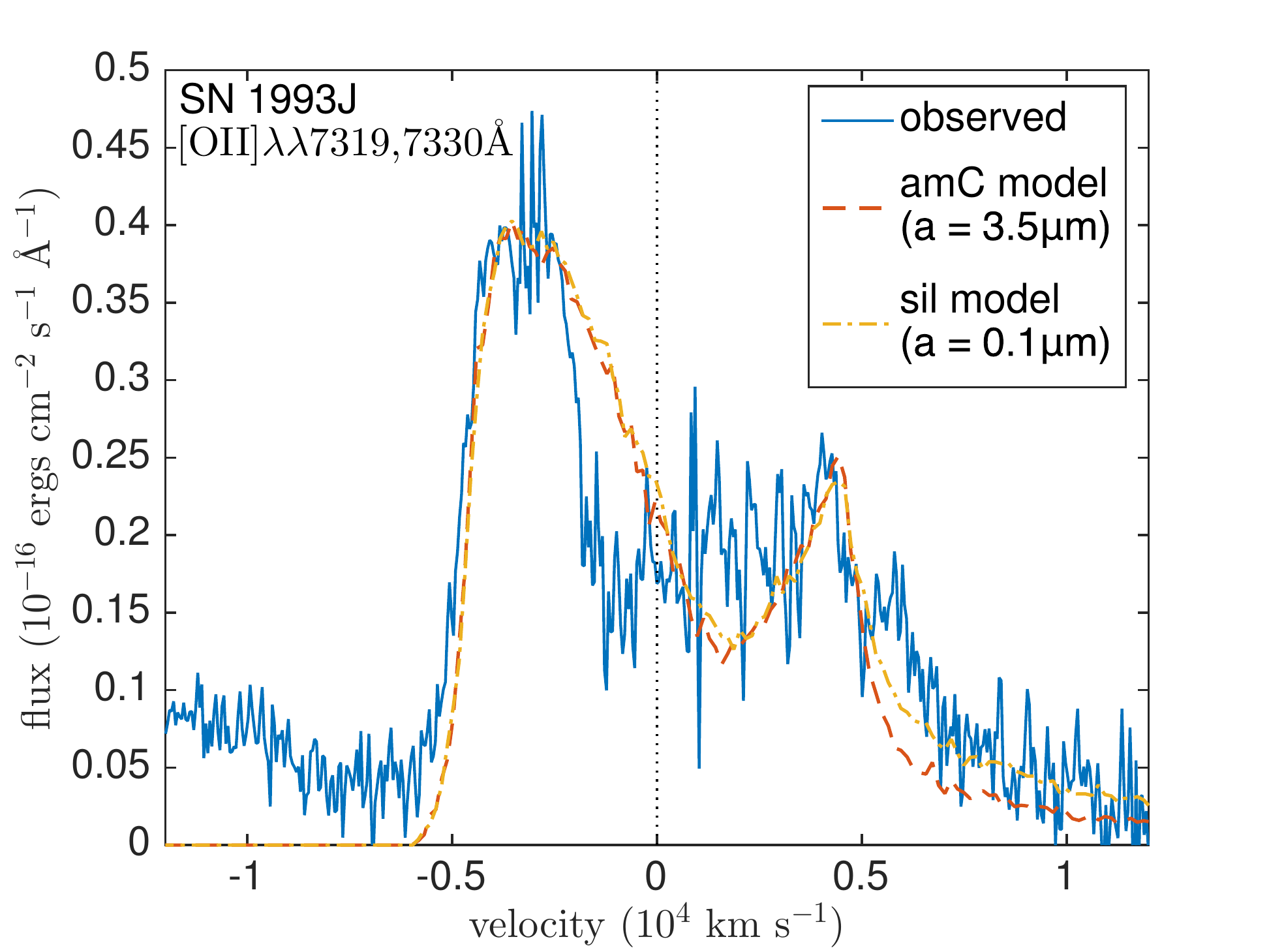} 
\caption{Best smooth dust fits to the  Year 16
SN~1993J [O~{\sc iii}] $\lambda\lambda$4959,5007 doublet ({\em top}) 
and the [O~{\sc ii}] $\lambda\lambda$7319,7330 doublet ({\em bottom}) 
for the parameters detailed in Table \ref{93J}.  For the [O~{\sc ii}] 
doublet, zero velocity was set at $\lambda=7319$~\AA\ and for the [O~{\sc 
iii}] doublet, zero velocity was set at $\lambda=4959$~\AA.  Compositions 
and grain sizes are as detailed on the plots.} 
\label{93J_smooth} 
\end{figure}

We had mixed success in obtaining good fits to the oxygen line profiles of 
SN~1993J.  Certain aspects of the line profiles are well-fitted by the 
models, such as the bump seen at $-4000$~km~s$^{-1}$ in the frame of
[O~{\sc iii}] $\lambda4959$. The peaks of the [O~{\sc iii}] profile were 
fairly well matched 
although we could not exactly reproduce the peak seen at $+4000$~km~s$^{-1}$
(Figures \ref{93J_smooth} and \ref{93J_clumped}).
Similarly, we struggled to reproduce the shape of the profile between 
$-2000$ and $+2000$~km~s$^{-1}$ for both the [O~{\sc iii}] and [O~{\sc 
ii}] doublets.  In order to fit the profile, we needed to use different 
sized grains for the [O~{\sc iii}] and the [O~{\sc ii}] models in order to 
fit the red wings of the profiles. Parameters for the smooth dust models 
for [O~{\sc iii}] $\lambda\lambda$4959,5007 and [O~{\sc ii}]
$\lambda\lambda$7319,7330 that are presented in Figure 
\ref{93J_smooth} are detailed in Table \ref{93J}.

\subsection{SN~1993J Clumped Dust Models}

\begin{figure} 
\centering 
\includegraphics[scale=0.38,clip=true, trim=10 0 40 20]{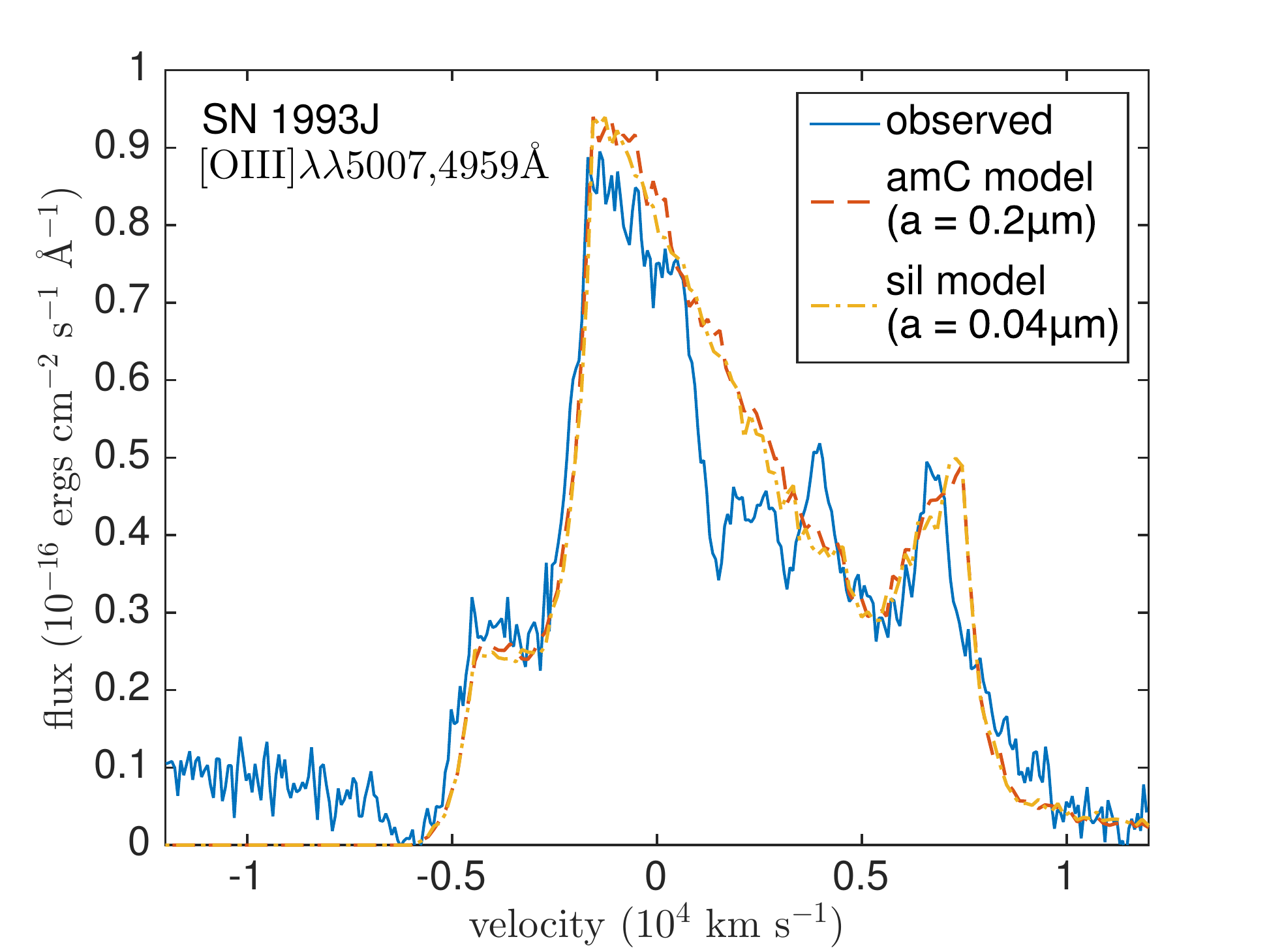}
\includegraphics[scale=0.38,clip=true, trim=0 0 40 20]{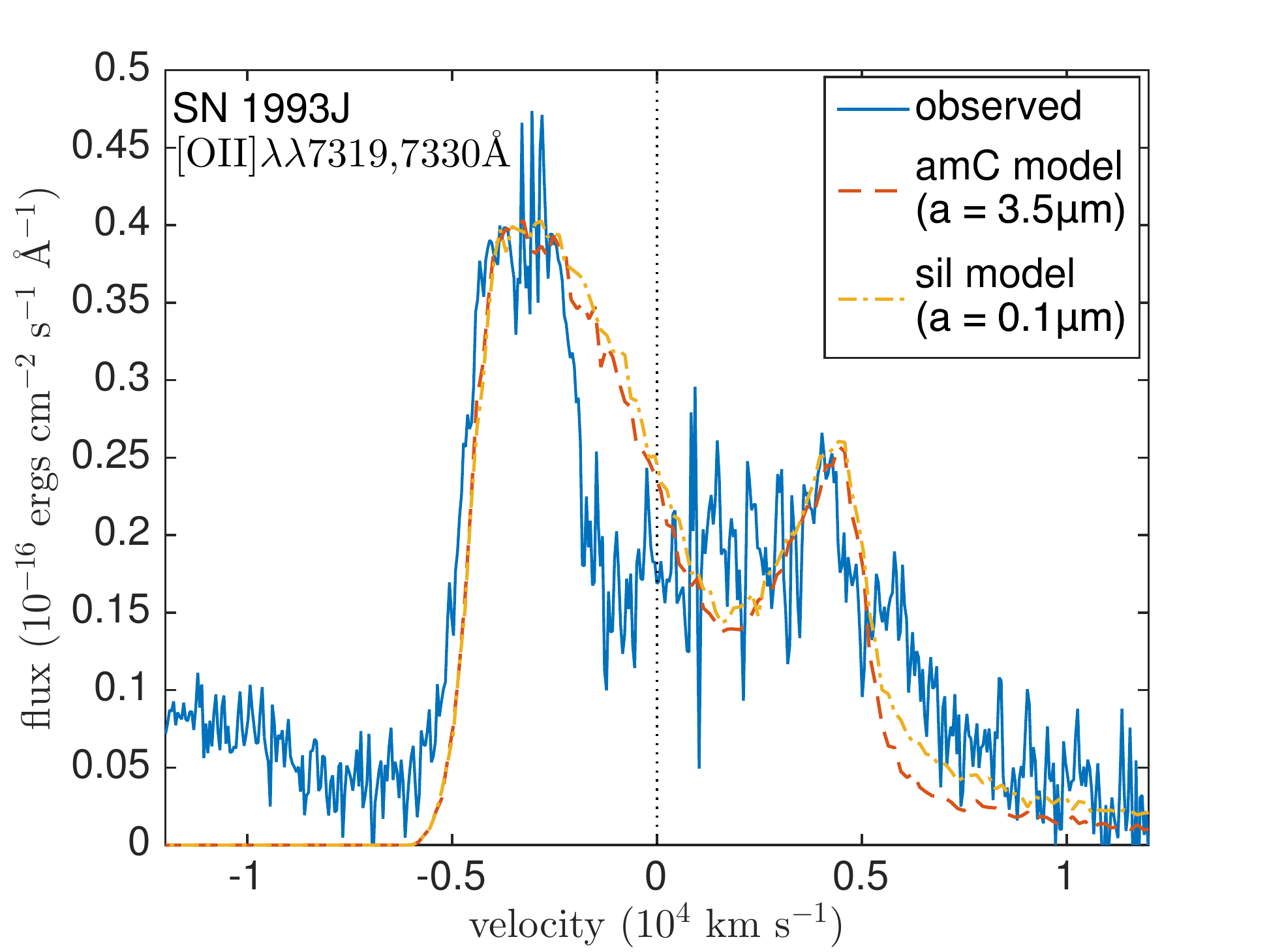} 
\caption{Best clumped dust fits to the Year 16
SN~1993J [O~{\sc iii}] $\lambda\lambda$4959,5007 doublet ({\em 
top}) and the [O~{\sc ii}] $\lambda\lambda$7319,7330 doublet ({\em 
bottom}) for the parameters detailed in Table \ref{93J}.  For the [O~{\sc 
ii}] doublet, zero velocity was set at $\lambda=7319$~\AA\ and for the 
[O~{\sc iii}] doublet, zero velocity was set at $\lambda=4959$~\AA. 
Compositions and grain sizes are as detailed on the plots.} 
\label{93J_clumped} 
\end{figure}

Our clumped dust models for SN~1993J adopt the same clumped structure as 
for the SN~1980K clumped models and the SN~1987A clumped models ($f=0.1$ 
and $R_{clump}=R_{out}/25$).  All of the parameters were kept fixed from 
the smooth models except for the new clumped dust distribution.  Similar 
fits were found and the clumped geometry had little effect on the 
resultant modelled line profiles.  The required dust mass increased by a 
factor of approximately 1.5.  The clumped model parameters for the 
[O~{\sc iii}] $\lambda\lambda$4959,5007 and [O~{\sc ii}]
$\lambda\lambda$7319,7330 fits that are presented in Figure 
\ref{93J_clumped} are detailed in Table \ref{93J}.

\subsection{SN~1993J Discussion} 

Our models for the optical line profiles of SN~1993J do not fit the 
observed data quite as well as the models for the other objects.  In 
particular, the modelled profiles tend to over-estimate the flux in the 
region just to the red side of the peak flux, where the observed profile 
exhibits a sharp downturn (see Figure \ref{boxy}).  The steepness of this 
drop cannot be matched by the models.  However, certain other features of 
the observed profiles are fitted well by the model.  For example, the 
[O~{\sc iii}] $\lambda\lambda$4959,5007 model line profile in particular 
fits bumpy features on both the red and blue sides of the profile at 
approximately $-4000$~km~s$^{-1}$ and $+6000$~km~s$^{-1}$ quite well.  
The bump near $+6000$~km~s$^{-1}$ is simply due to absorption in the 
region between $-V_{min}$ and $+V_{min}$ causing a peak at the location of 
the minimum velocity (for the 5007~\AA\ component).  The small discrepancy 
in the location of the [O~{\sc iii}] peak at around 
$\sim+6000$~km~s$^{-1}$ may be a result of a net velocity shift of the 
supernova away from the observer (see Section \ref{scn:CasA_smooth} for a 
discussion of such effects in more detail) or a discrepancy between the 
adopted smooth, symmetrical gaseous emission models and a more clumpy, 
asymmetrical geometrical structure for the remnant \citep{Tran1997}.  
Similarly, the difference in grain sizes, and hence dust masses, required 
to fit the [O~{\sc iii}] and [O~{\sc ii}] lines (see Table \ref{93J}) may 
also indicate the need for a different distribution of dust or gas.  The 
bumpy features seen in these lines were discussed by \citet{Matheson2000b} 
who postulated that the `double-horned' shape was a consequence of the 
ejecta colliding with a disk-like or flattened region. While it is 
possible that the blue-shifted asymmetry observed in the optical line 
profiles is not a result of dust in the ejecta, given how well certain 
aspects of the observed profiles are fitted, it seems more likely that it 
is simply the case that a more complex emission geometry is required for 
the models to better fit the data.

A manual investigation of parameter space suggested that the issues with 
fitting the shapes of the profiles derived from the initial emissivity 
distribution.  The smooth nature of the emissivity distribution that we 
adopted could not reproduce the sharp downturn seen to the red side of the 
peak flux.  This could be due to an oxygen emission distribution that is 
potentially composed of a dense central region that produces the steep 
variations in the central regions, with a more diffuse oxygen envelope 
accounting for the wings.  

\citet{Nozawa2010} have presented post-explosion elemental abundance 
profiles computed for an analogue of a Type IIb supernova (such as Cas~A 
or SN~1993J), with an initial mass of 18~M$_\odot$ and a residual 
stellar mass of 4.5~M$_\odot$ at the time of explosion. Oxygen was 
predicted to be the dominant element in the inner 1.1~M$_\odot$ of the 
2.9~M$_\odot$ ejecta, and to have a significant abundance in the 
1.3~M$_\odot$ layer above that (O/He$\sim3\times10^{-3}$, but with a 
C/O ratio of 
$\sim$2.5). Our overestimation of the flux just to the red side of the 
peak might be resolved by considering a two-component density-velocity 
distribution such as this. SN~1993J had a particularly unusual red 
supergiant progenitor with a postulated stripped envelope caused by the 
presence of a B-star binary companion \citep{Maund2004,Fox2014}
resulting in a significant mass of circumstellar material surrounding the 
progenitor star.  Photometric analyses performed by \citet{Zhang2004} suggested 
that the late-time optical emission from SN~1993J is largely powered by 
interaction between the blast wave and the circumstellar material.  In 
this case, the geometry of the emitting regions is especially complex, and 
may in particular account for the significant substructure seen in the 
optical line profiles.


For both SN~1980K and SN~1993J, the effects of including a clumped dust 
distribution in the models rather than a smooth dust distribution 
increased the required dust mass by a factor of approximately 1.5, very 
similar to the factor found for SN~1987A from the models presented by 
\citet{Bevan2016}. In these cases, the clumped geometry has little effect 
on the resulting profiles except to reduce the degree of absorption.  The 
extent of the extended red scattering wing is also somewhat reduced by 
dust clumping. Since SN~1993J's spectrum is dominated by oxygen lines,
we prefer our silicate dust models to those that used amorphous carbon, 
implying a clumped dust mass of 0.08~--~0.15~M$_\odot$ for SN~1993J at 
year 16 (Table \ref{93J}).

\section{Cassiopeia A} 
\label{CasA_intro}

Cassiopeia~A (Cas~A) is a young supernova remnant at an estimated distance 
of 3.4~kpc \citep{Reed1995}. The X-ray 
radii of the reverse and forward shocks measured by
{\em Chandra} were  95$\pm$10~arcsec and 
153$\pm$12~arcsec \citep{Gotthelf2001}, corresponding to physical radii of 
1.57 and 2.52~pc respectively. Analysis of its expansion velocities and 
geometry have allowed the explosion date to be estimated as
1681$\pm$19~CE \citep{Fesen2006} implying that the remnant 
was approximately 330 years old at the time that the spectra discussed 
here were obtained. Cas~A is the strongest radio source in the sky outside 
of the solar system and has been comprehensively observed across the 
electromagnetic spectrum.  Optical spectroscopy of light echoes from the 
supernova scattered off surrounding interstellar dust has allowed the 
original supernova explosion to be spectroscopically classified, leading 
to the conclusion that Cas~A was the result of a Type IIb supernova 
explosion \citep{Krause2008}.

\citet{Arendt1999} estimated a mass of 0.038~M$_{\odot}$ of 52~K dust in 
Cas~A based on fitting IRAS 60~$\mu$m and 100~$\mu$m fluxes. From {\em 
Spitzer} IRS spectroscopy and MIPS photometry \citet{Rho2008} estimated 
that a mass of 0.020~--~0.054~M$_{\odot}$ of warm 65~--~265~K dust was 
emitting between 5~$\mu$m and 70~$\mu$m, particularly in a bright ring 
associated with the reverse shock. \citet{Arendt2014} used {\em Spitzer} 
IRS spectra and PACS 70~--~160-$\mu$m photometry to estimate a mass of 
$\sim$0.04~M$_{\odot}$ of warm dust in Cas~A.

Observations at longer wavelengths of the cold dust in the ejecta have led 
to higher dust mass estimates, with \citet{Dunne2003} using SCUBA 450- and 
850-$\mu$m photometry to estimate a dust mass of 2~--~4~M$_{\odot}$.  This 
was contested by \citet{Krause2004} who suggested that the majority of the 
dust emission originated from cold interstellar clouds located along the 
line of sight to Cas~A and placed an upper limit of 0.2~M$_{\odot}$ of 
cold dust in the ejecta.  However, observations of strongly polarised 
850-$\mu$m emission obtained using the SCUBA polarimeter have been used to 
argue for the presence of an ejecta-condensed cold dust mass of 
$\sim0.9$~M$_{\odot}$ \citep{Dunne2009}.

Modelling by \citet{Nozawa2010} reproduced the observed {\em IRAS}, {\em 
ISO} and {\em Spitzer} infrared SED using 0.08~M$_{\odot}$ of dust, of 
which 0.072~M$_{\odot}$ was inside the radius of the reverse shock.  
\citet{Sibthorpe2010} used far-infrared and sub-mm {\em AKARI} and 
balloon-borne BLAST observations to estimate a 35~K dust mass of 
0.06~M$_{\odot}$ in Cas~A, while \citet{Barlow2010} using higher angular 
resolution {\em Herschel} PACS and SPIRE photometry to estimate a mass of 
0.075$\pm$0.028~M$_{\odot}$ of T$\sim35$~K silicate dust located mainly 
inside the reverse shock, which combined with the previously estimated 
mass of 0.02~--~0.05~M$_{\odot}$ of warmer dust associated with the reverse 
shock yielded a total dust mass estimate of $\sim$0.1~M$_{\odot}$ for 
Cas~A.

\subsection{The Integrated Optical Spectrum of Cas~A} \label{CasA_integr}

\begin{figure*} 
\centering 
\includegraphics[clip=true,scale=0.6, trim=30 0 50 20]{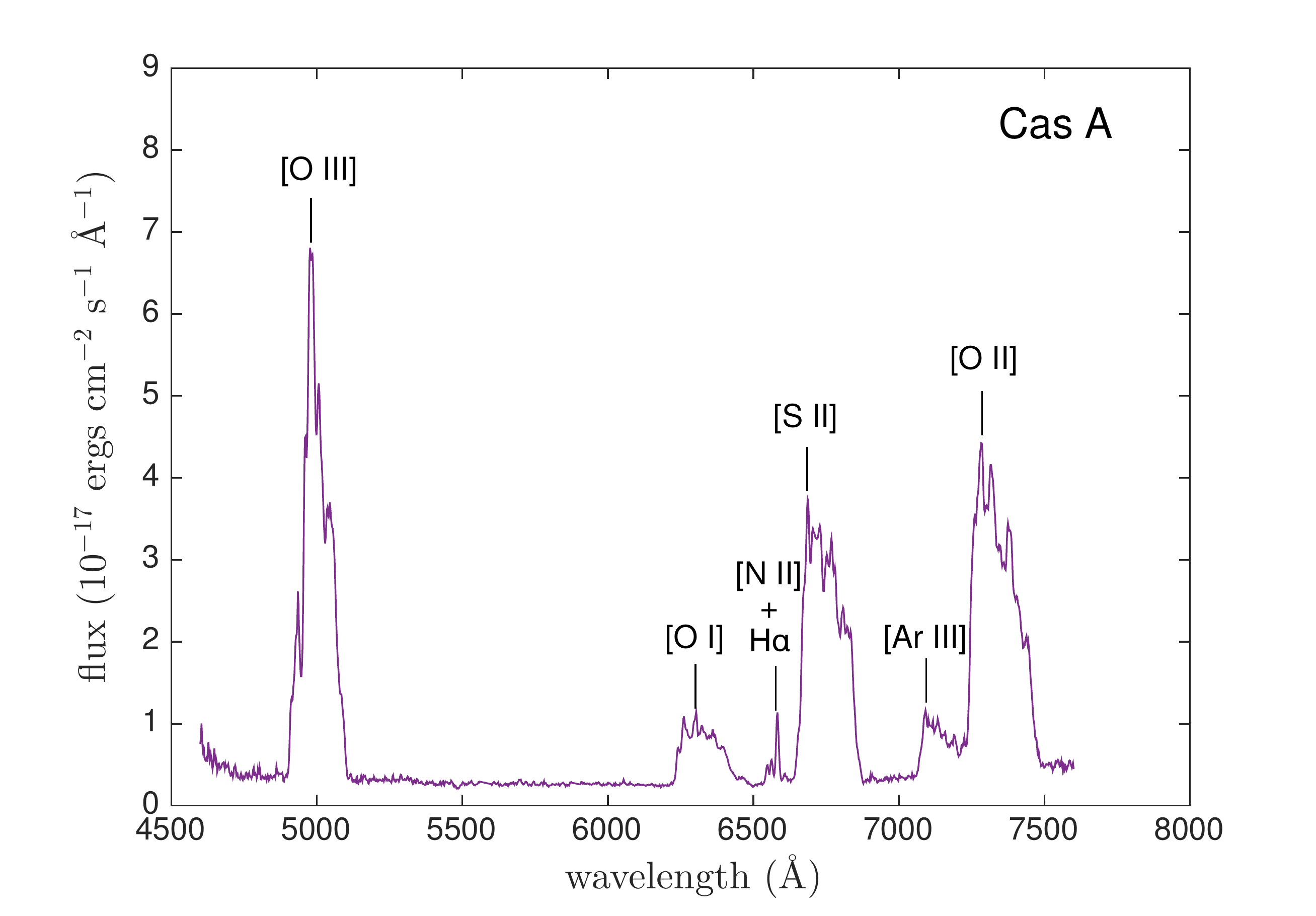} 
\caption{The integrated spectrum of Cas~A - see 
\citet{Milisavljevic2013} and Section \ref{CasA_integr}}. 
\label{CasA_spectrum} 
\end{figure*}

The integrated optical spectrum of Cas~A, obtained by 
\citet{Milisavljevic2013} and discussed by \citet{Milisavljevic2012}, is
displayed in Figure \ref{CasA_spectrum}. It shows red-blue asymmetries in 
many of the line profiles. In particular, the oxygen lines [O~{\sc i}] 
$\lambda\lambda$6300,6363, [O~{\sc ii}] $\lambda\lambda$7319,7330~ and 
[O~{\sc iii}] $\lambda\lambda$4959,5007 exhibit a blue-shifted asymmetry, 
with the [O~{\sc iii}] doublet especially demonstrating a strong blueshift 
with considerable substructure.  We have modelled all three of these 
features with a primary focus on the [O~{\sc iii}] doublet.

The integrated spectrum was composed of observations carried out mainly in 
September 2007 and September 2008 at Kitt Peak, Arizona using the MDM 2.4m 
Hiltner telescope, as described by \citep{Milisavljevic2013}.  The MDM 
Modular Spectrograph was used with a 2048$\times$2048 CCD detector and a
long slit of dimensions 2$''\times5'$ oriented North-South.  
Exposure times were generally 2$\times$500s. The wavelength range covered 
was 4500~--~7000~\AA\ with a spectral resolution of 6~\AA. The integrated 
spectrum was ultimately composed of 80 long slit spectra spaced 3$''$ 
apart across the entire main shell. The slit positions are shown in 
Figure~1 of \citet{Milisavljevic2013}.

\subsection{Smooth Dust Models for the Oxygen Lines of Cas A} 
\label{scn:CasA_smooth}

The modelling of the Cas~A spectrum was initially focussed on the [O~{\sc 
iii}] $\lambda\lambda$4959,5007 doublet, which exhibits a pronounced 
asymmetry. The process of finding a fit to the line profile was the same 
as described earlier. The maximum velocity was identified from 
the point at which flux vanishes on the blue side, the inner to outer 
radius ratio was determined from various inflection points and the density 
profile was determined from the shape of the profile.  The other 
parameters were then iterated to find the best fitting profile.


We initially produced a fit to the data using the parameters 
listed in the first row of Table \ref{CasA_smooth_params}.  The profile is 
presented in the top pane of Figure \ref{CasA_OIII}.  As can be seen, the 
modelled line profile fits most of the observed line profile, 
but fails to fit the red side of the profile adequately.  A 
thorough, manual investigation of parameter space resulted in the 
conclusion that the profile was much better fitted if the entire modelled 
profile was shifted to the red by $+700$~km~s$^{-1}$. There is other 
evidence in favour of this assumption. The dynamics of Cas~A are known to 
be significantly 
asymmetrical \citep{Rest2011}, with radial velocities spanning -4000 to 
+6000~km~s$^{-1}$ \citep{Milisavljevic2013} suggesting that the net 
line-of-sight velocity is likely away from the observer and indicating the 
need for an overall velocity shift to correct for this.  Using {\em 
Spitzer} IRS 5~--~40-$\mu$m spectra to construct a 3D model of the remnant, 
\citet{DeLaney2010} derived an average velocity offset away from the 
observer of +859~km~s$^{-1}$. From their optical reconstruction of
the remnant,  \citet{Milisavljevic2013} derived an offset of 
+760~$\pm$~100~km~s$^{-1}$ that is consistent with our own estimate.

We found that models of the [O~{\sc ii}] $\lambda\lambda$7319,7330 and 
[O~{\sc i}] $\lambda\lambda$6300,6363 lines were also substantially 
improved if the entire model profile was allowed to be uniformly shifted 
towards the red.  For the remainder of the models we therefore shifted the 
profiles in velocity space to better fit the data based on the likelihood 
that the sampled emitting regions had an overall net velocity away from 
the observer.  Fits to the line profiles were significantly improved 
following this translation.

A model with the shifted [O~{\sc iii}] $\lambda\lambda$4959,5007 line is
presented in Figure \ref{CasA_OIII} for a dust
medium composed of 50\% amorphous carbon and 50\% silicate grains of
radius $a=0.05~\mu$m. The parameters used for this model 
are presented in the second row of Table \ref{CasA_smooth_params}.  A 
total dust optical depth of $\tau=0.49$ at 5007~\AA\ between $R_{in}$ and 
$R_{out}$ was found to best fit the profile.  An albedo of 
$\omega\approx0.15$ at 5007~\AA\ was needed in order to match the flux 
on the far red side of the model profile to that observed.

\afterpage{ 
\centering 
\setlength{\tabcolsep}{10pt} 
\begin{table*} 
\centering 
	\caption{The parameters used for the smooth models of Cas~A with a 
medium composed of 50\% amorphous carbon and 50\% silicate grains of 
radius $a=0.05~\mu$m. Dust optical depths are given from $R_{in}$ to 
$R_{out}$ at $\lambda = 5007$~\AA\ for [O~{\sc iii}], $\lambda = 7319$~\AA\ 
for [O~{\sc ii}] and $\lambda = 6300$~\AA\ for [O~{\sc i}]. The doublet ratio 
is always the ratio of the stronger line to the weaker line. The asterisk 
indicates that the parameters listed describe the gas density 
distribution.  The dust density distribution is the same in all cases (as 
detailed for the shifted [O~{\sc ii}] doublet in the second row).}
	\label{CasA_smooth_params}
	\centering
  	\begin{tabular}{@{} cccccccccccc @{}}
    	\hline
  &$v$ shift& $V_{max}$ & $V_{min}$ & $R_{in}/R_{out}$ & $\beta$ & 
$M_{dust}$ & $R_{out}$ & $R_{in}$ & doublet & $\tau_{\lambda}$ \\
	& (km~s$^{-1}) $& (km~s$^{-1} $)& (km~s$^{-1} $) & & & 
        ($M_{\odot}$) & (10$^{18}$cm) & (10$^{18}$cm) & ratio \\
\hline 
[O~{\sc iii}] & 0 & 4500 & 1800 & 0.4 & 2.0 & 0.9 & 4.7 & 1.9 & 2.98 & 0.53 \\ \relax 
[O~{\sc iii}] & +700 & 5000 & 2500 & 0.5 & 2.0 & 1.1 & 5.2 & 2.6 & 2.98 & 0.49 \\ \relax 
[O~{\sc ii}]* & +1000 & 5000 & 3250 & 0.65 & 2.0 & 1.1 & 5.2 & 3.4 & 1.23 & 0.21 \\ \relax 
[O~{\sc i}]* & +1000 & 5000 & 3250 & 0.65 & 2.0 & 1.1 & 5.2 & 3.4 & 3.1 & 0.30 \\
    \hline
  \end{tabular}

\end{table*} 

\begin{figure} 
\centering 
\includegraphics[scale=0.43,clip=true, trim=30 0 50 20]{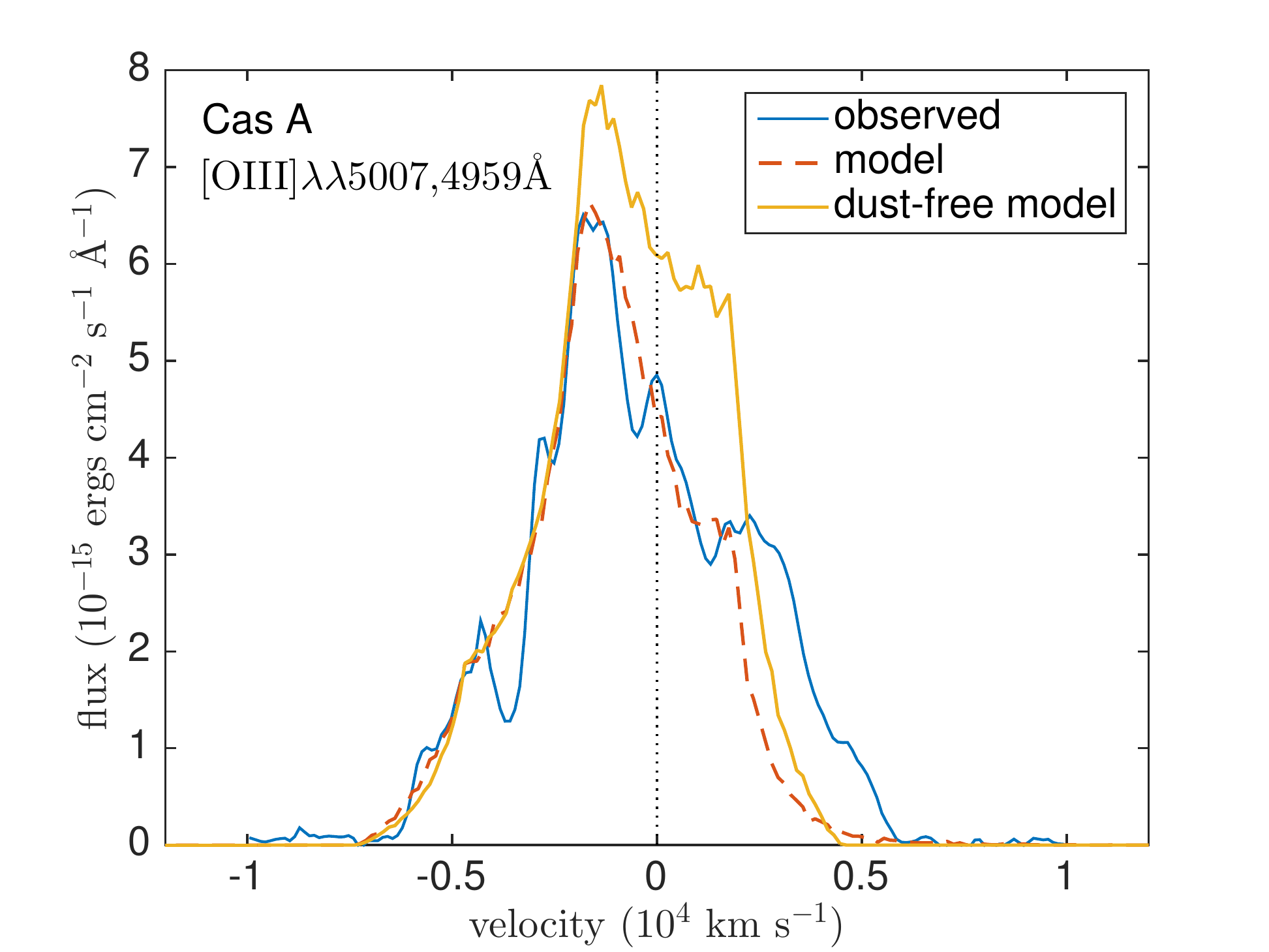} 
\includegraphics[scale=0.43,clip=true, trim=30 0 50 20]{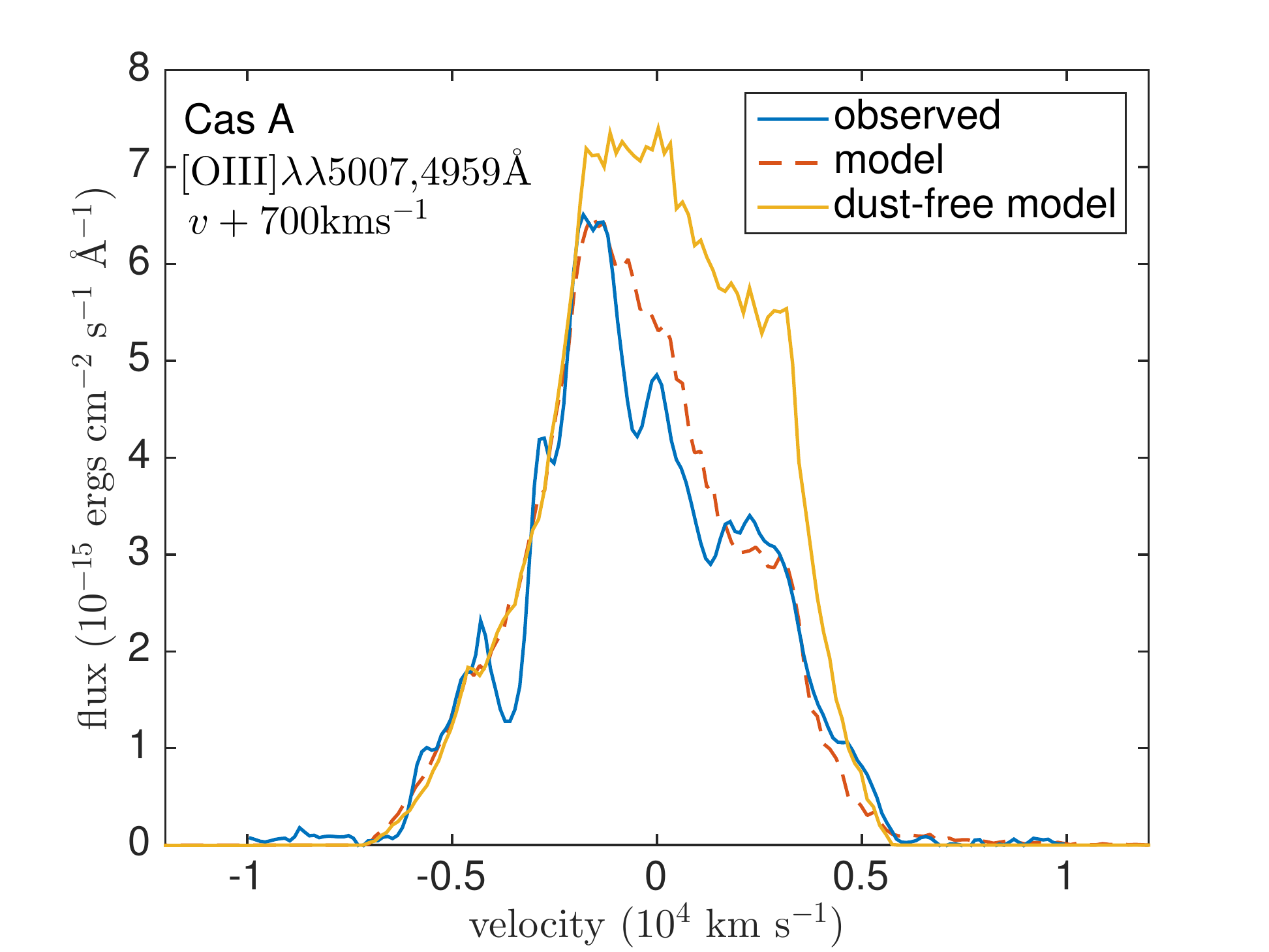} 
\caption{Best smooth dust fits to the Cas~A [O~{\sc iii}] 
$\lambda\lambda$4959,5007 doublet for the parameters detailed in Table 
\ref{CasA_smooth_params}, along with the modelled intrinsic line
profile (yellow). The top plot shows an unshifted fit to the 
[O~{\sc iii}] line profile and in the bottom plot the model 
[O~{\sc iii}] line has been shifted uniformly towards the red by 
$+700$~km~s$^{-1}$. Zero velocity was set at $\lambda=4959$~\AA.}
\label{CasA_OIII} 
\end{figure} 
}

The composition of the dust has a significant effect on the overall dust 
mass for this optical depth and albedo. An attenuated line 
profile model of the [O~{\sc iii}] $\lambda\lambda$4959,5007 doublet from 
Cas~A could not be found using 100\% astronomical silicate dust 
\citep{Draine1984}.  There is little to no red scattering wing seen, hence 
the relatively low value of $\omega$, and therefore relatively small 
silicate grains would be required to reproduce the red side of the 
profile.  Silicate grains of this size have extremely low optical 
absorption efficiencies and therefore the best-fitting optical depth of 
$\tau=0.49$ would correspond to an implausibly large mass of dust 
($>20$~M$_{\odot}$) if it was composed entirely of astronomical silicates.

The chemical composition of the dust in the ejecta of Cas~A is known to be 
extremely complex \citep{Rho2008,Arendt2014} with many different species 
of dust grain present in the ejecta.  The presence of silicate dust has 
been deduced based on the silicaceous emission features observed in the 
mid-IR region of the spectrum \citep{Rho2008}.  However, the potential
presence of a variety of other species has been discussed 
\citep{Arendt2014}. In Table \ref{CasA_dust_masses}, we detail the dust 
masses required to fit the [O~{\sc iii}]$\lambda\lambda$4959,5007 
line profile for different fractions of silicates and amorphous carbon 
grains for a single grain size.  For each composition we determined the 
grain radius based on the albedo necessary to fit the profile 
($\omega\approx0.15$) and then varied the dust mass to achieve the 
required optical depth. The derived dust masses cover a wide range of 
values, between 0.37~--~6.5~M$_{\odot}$.

We investigated whether it was possible to determine the approximate dust 
composition based on the relative optical depths necessary to fit 
different blue-shifted lines in the spectrum and the wavelength dependence 
of dust absorption for different grain compositions and sizes. We 
therefore considered the 
blue-shifted [O~{\sc ii}] $\lambda\lambda$7319,7330 and [O~{\sc iii}] 
$\lambda\lambda$4959,5007~\AA\ lines from Cas~A.  Unfortunately, at the 
small grain sizes required, there is no significant variation between the 
absorption efficiencies at 5007~\AA\ and 7319~\AA\ for different dust 
compositions and we could not therefore determine the composition via this 
approach.  Additionally, the [O~{\sc ii}] and [O~{\sc i}] lines are much 
less sensitive to variations in the density distributions and the dust 
mass, partly due to the high frequency of bumpy features observed in these 
lines which contaminate the intrinsic broad profile.  The best-fitting 
models for these lines were therefore quite degenerate i.e. there were 
multiple sets of parameters that resulted in reasonable fits.

\begin{figure} 
\centering 
\includegraphics[scale=0.41,clip=true, trim=15 0 40 20]{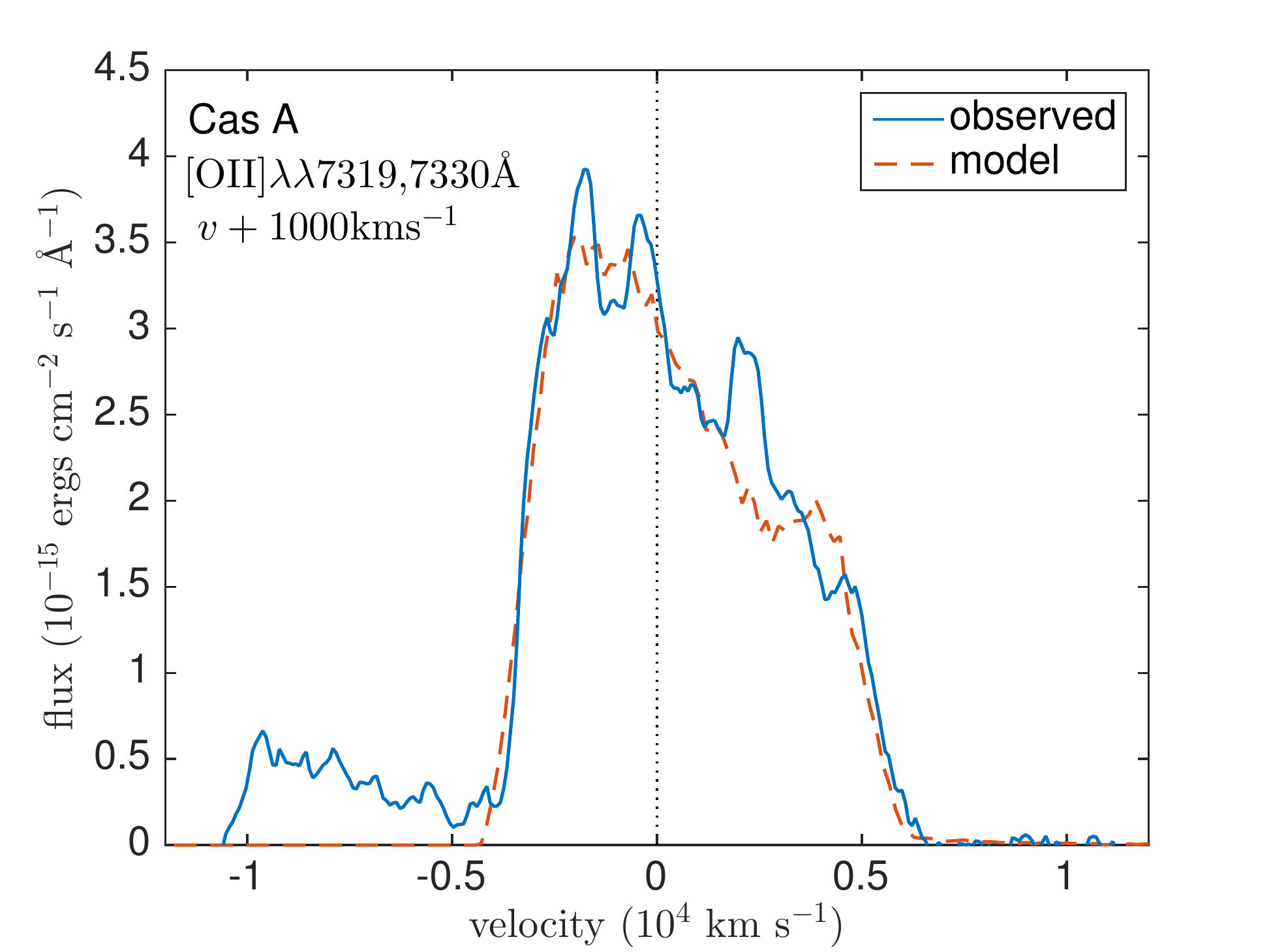} 
\includegraphics[scale=0.41,clip=true, trim=15 0 40 20]{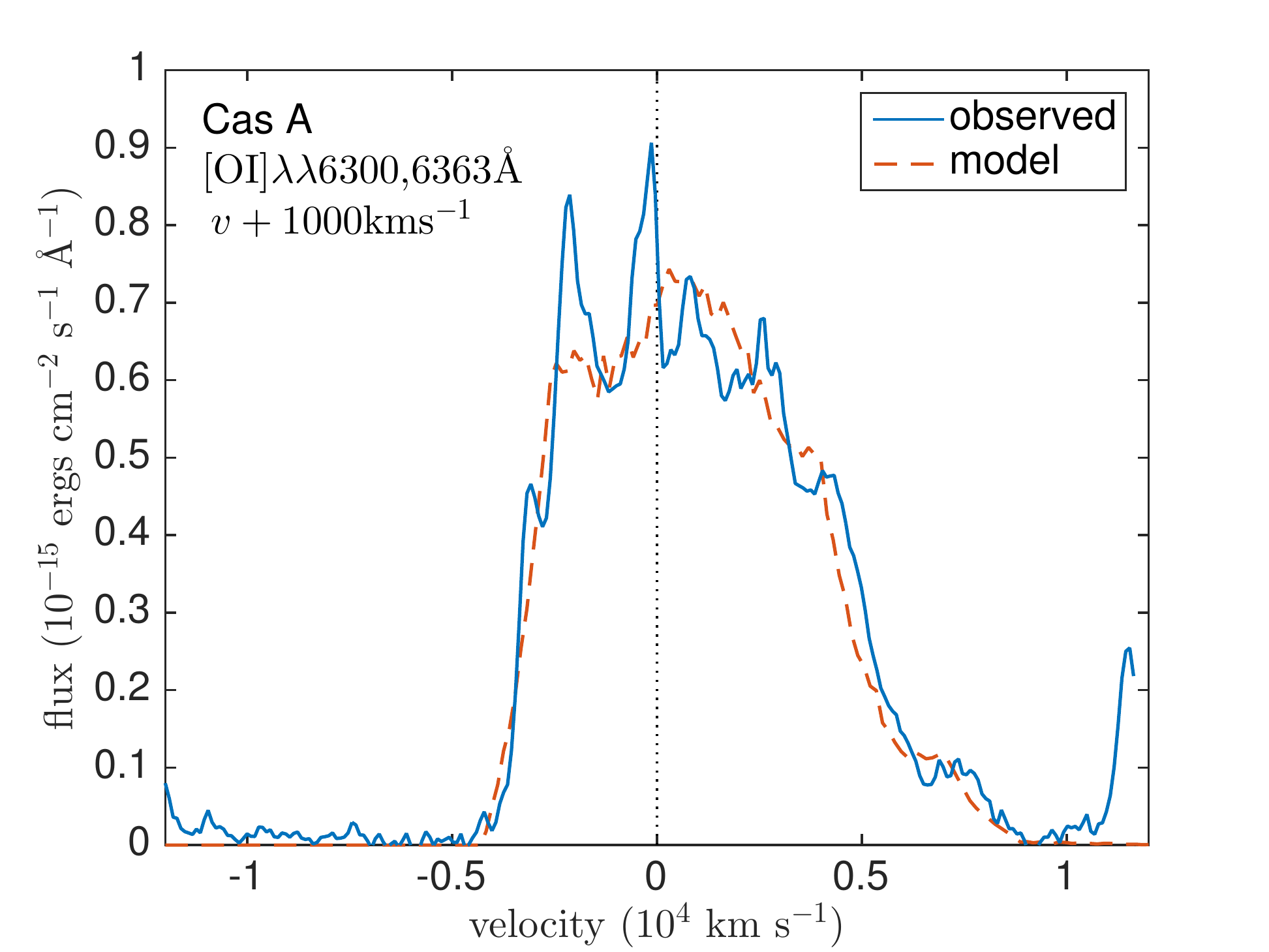} 
\caption{Best smooth dust fits to the 
Cas~A [O~{\sc ii}] $\lambda\lambda$7319,7330 doublet ({\em top}) and 
the [O~{\sc i}] $\lambda\lambda$6300,6363 doublet ({\em bottom}) for 
the parameters detailed in Table \ref{CasA_smooth_params}. Both model 
line profiles have been shifted uniformly towards the red by 
$+1000$~km~s$^{-1}$.} 
\label{CasA_OI_OII} 
\end{figure}

However, it was possible to use these lines to determine the reliability 
of the best-fitting model for the [O~{\sc iii}] $\lambda\lambda$4959,5007 
line profile. We adopted the dust distribution determined using the 
[O~{\sc iii}] fits and investigated models for the [O~{\sc ii}] 
$\lambda\lambda$7319,7300 and [O~{\sc i}] $\lambda\lambda$6300,6363 
profiles to see if this dust distribution was capable of fitting these 
lines as well. We adopted an emissivity distribution that was slightly 
different to the [O~{\sc iii}] line (see Table \ref{CasA_smooth_params}) 
and shifted the observed line profiles by $-1000$~km~s$^{-1}$.  These 
emissivity distributions were then modelled with the dust distribution and 
dust mass for the best-fitting smooth [O~{\sc iii}] model.  The resultant 
[O~{\sc ii}] and [O~{\sc i}] line profiles can be considered good fits 
(see Figure \ref{CasA_OI_OII}). This suggests that the models are 
consistent and, if the relative abundances of the dust grain species 
present in the ejecta can be determined via other means, that the dust 
mass can be well-constrained using this method.  As in the cases of the 
line profile fitting for SN~1980K and SN~1993J, all of the line profile 
models listed above adopted intrinsic doublet strengths from 
\citet{Zeippen1987} and \citet{Storey2000}.

\subsection{Clumped Dust Models for the Oxygen Lines of Cas A}

Cas~A is highly clumped \citep{Fesen2001}. Recently, models 
by \citet{Biscaro2014} have suggested that dust cannot in fact form in the 
gas phase in the ejecta of Cas~A unless extremely dense knots of material 
are present.  It is therefore important, as with SN~1987A, to consider the 
effects of dust clumping on the line profiles. We continue to focus on the 
[O~{\sc iii}] line profile from Cas~A in considering the effects of 
clumping.  Clearly, the ejecta have a complex geometry with many clumps of 
different sizes and potentially different ionisation states and dust 
species 
within each.  The models that we present here are included to give some 
indication of the effects of clumping within the ejecta rather than to be 
representative of the state of the ejecta at this time.  To this end we 
present a number of models of the [O~{\sc iii}] line profile based on the 
smooth dust fits that we presented in the previous section. We consider 
two different clump sizes, ones with width $R_{out}/25$ and ones with 
width $R_{out}/10$. We also consider three different clump volume filling 
factors $f=0.05$, $f=0.1$ and $f=0.25$.  For each combination of clump 
size and filling factor we evaluated the required increase or decrease in 
the dust mass over the smooth dust model.  All other parameters were kept 
fixed so that line packets were emitted according to 
the smooth distribution and geometry described by the parameters listed 
in Table \ref{CasA_smooth_params}.

\begin{table} 
\centering 
	\caption{The variation in dust mass for a fixed dust optical depth 
$\tau_{5007}=0.49$ for the smooth dust parameters listed in Table 
\ref{CasA_smooth_params}.}
	\label{CasA_dust_masses}
	\centering
  	\begin{tabular}{c c c c}
    	\hline
	\% silicate & \% amorphous & grain radius $a$ & $M_{dust}$ \\
	grains& carbon grains&($\mu$m)&(M$_{\odot}$)\\
		\hline 
90 &10 &0.035& 6.5 \\ 
75 &25 &0.04 &2.5\\ 
50 &50 &0.045& 1.1\\ 
25 &75 &0.048& 0.6\\ 
0 &100 &0.05& 0.37\\
    \hline
  \end{tabular} 
  \end{table}

\begin{figure*} 
\centering 
\includegraphics[scale=0.43,clip=true, trim=30 0 50 20]{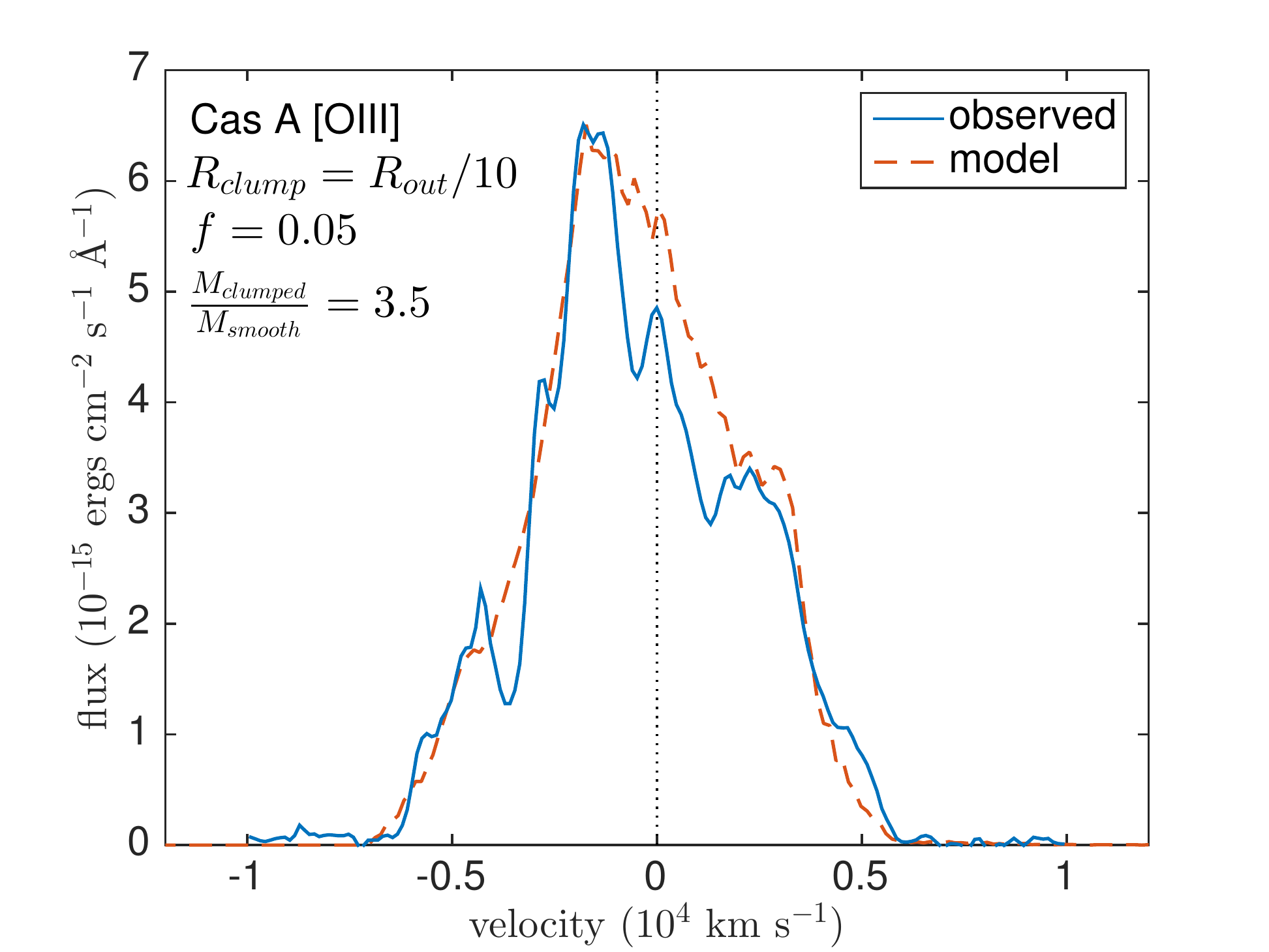} 
\includegraphics[scale=0.43,clip=true, trim=30 0 40 20]{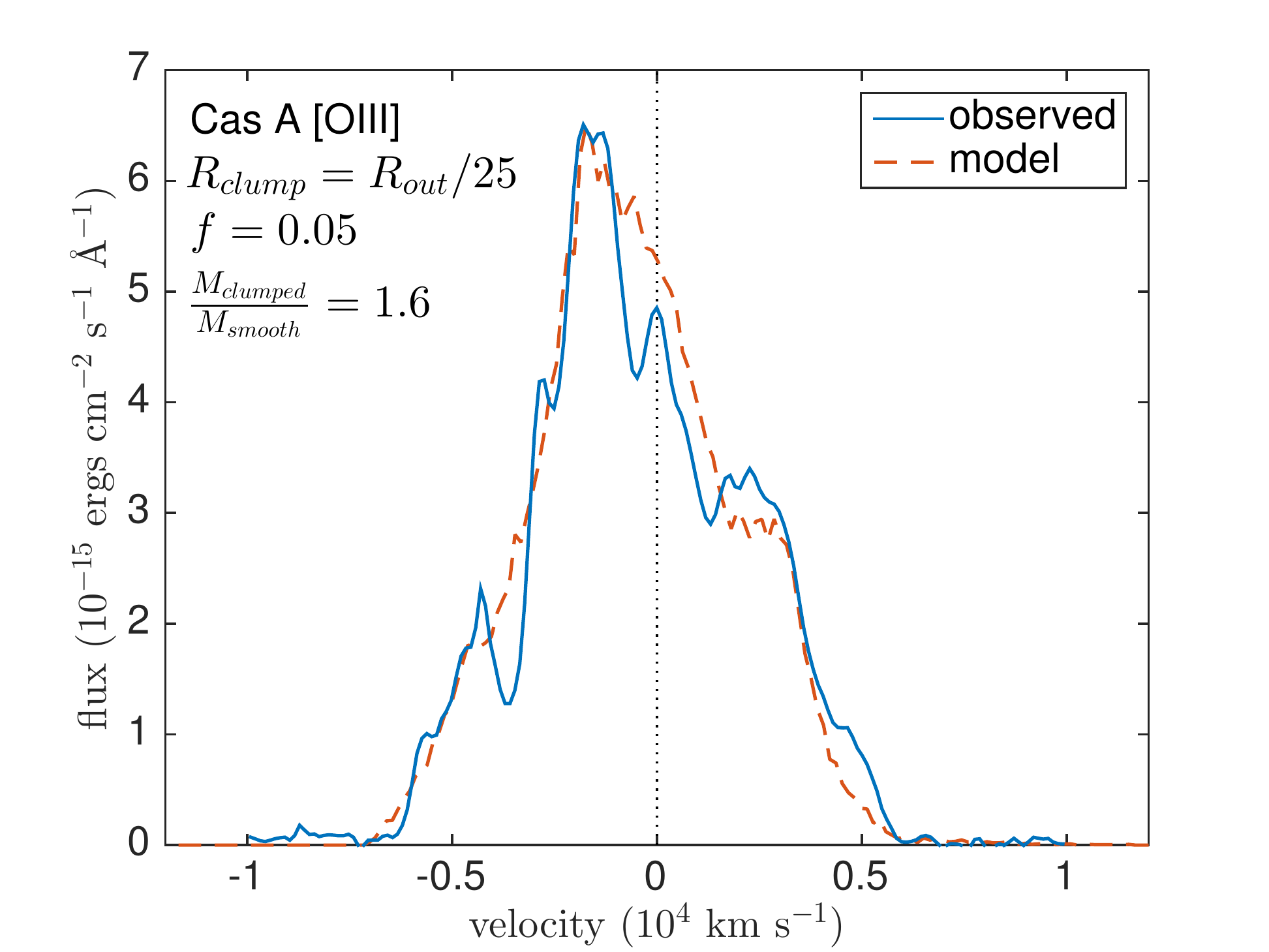}

\vspace{6mm} 
\includegraphics[scale=0.43,clip=true, trim=30 0 50 20]{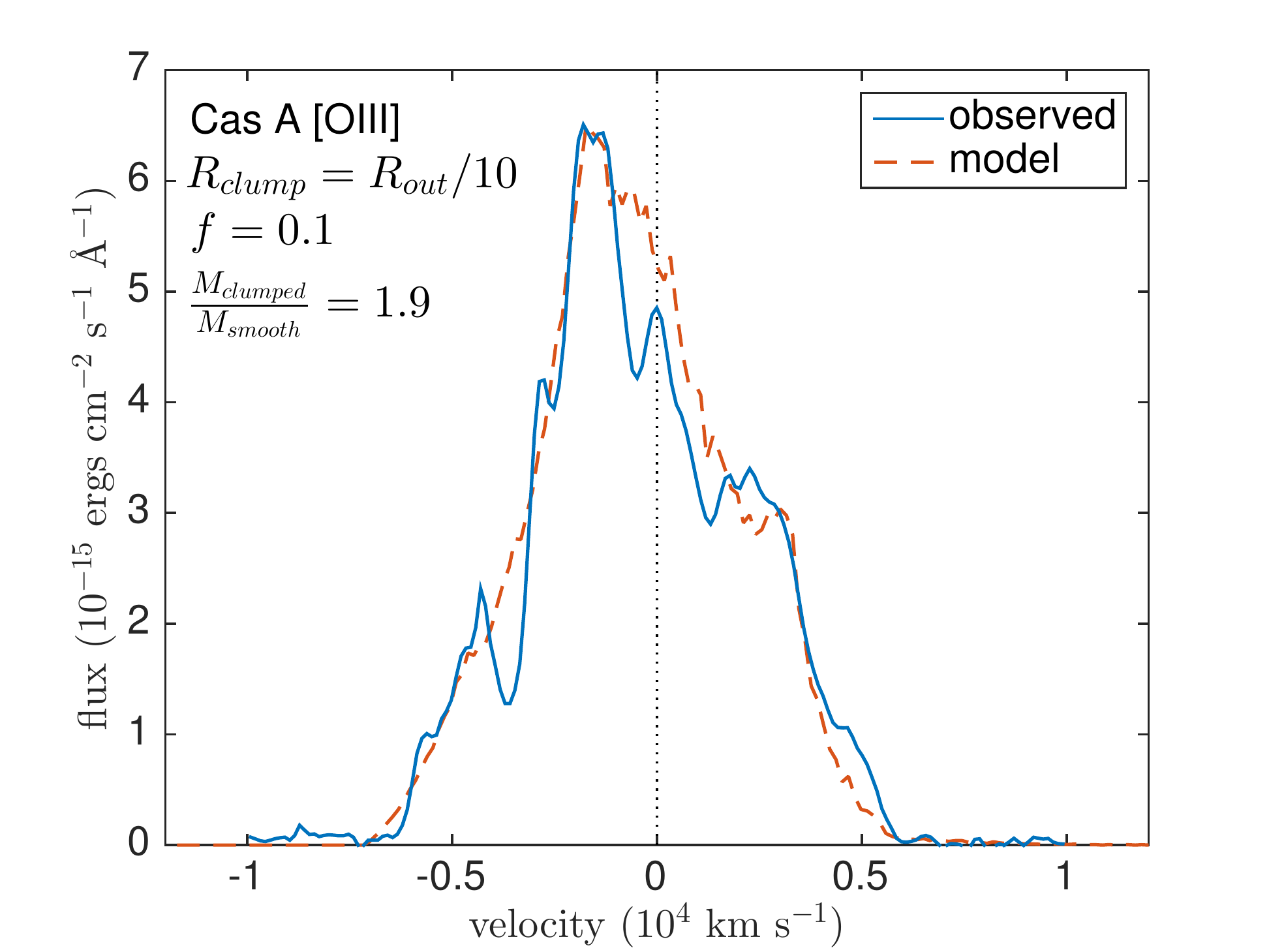} 
\includegraphics[scale=0.43,clip=true, trim=30 0 40 20]{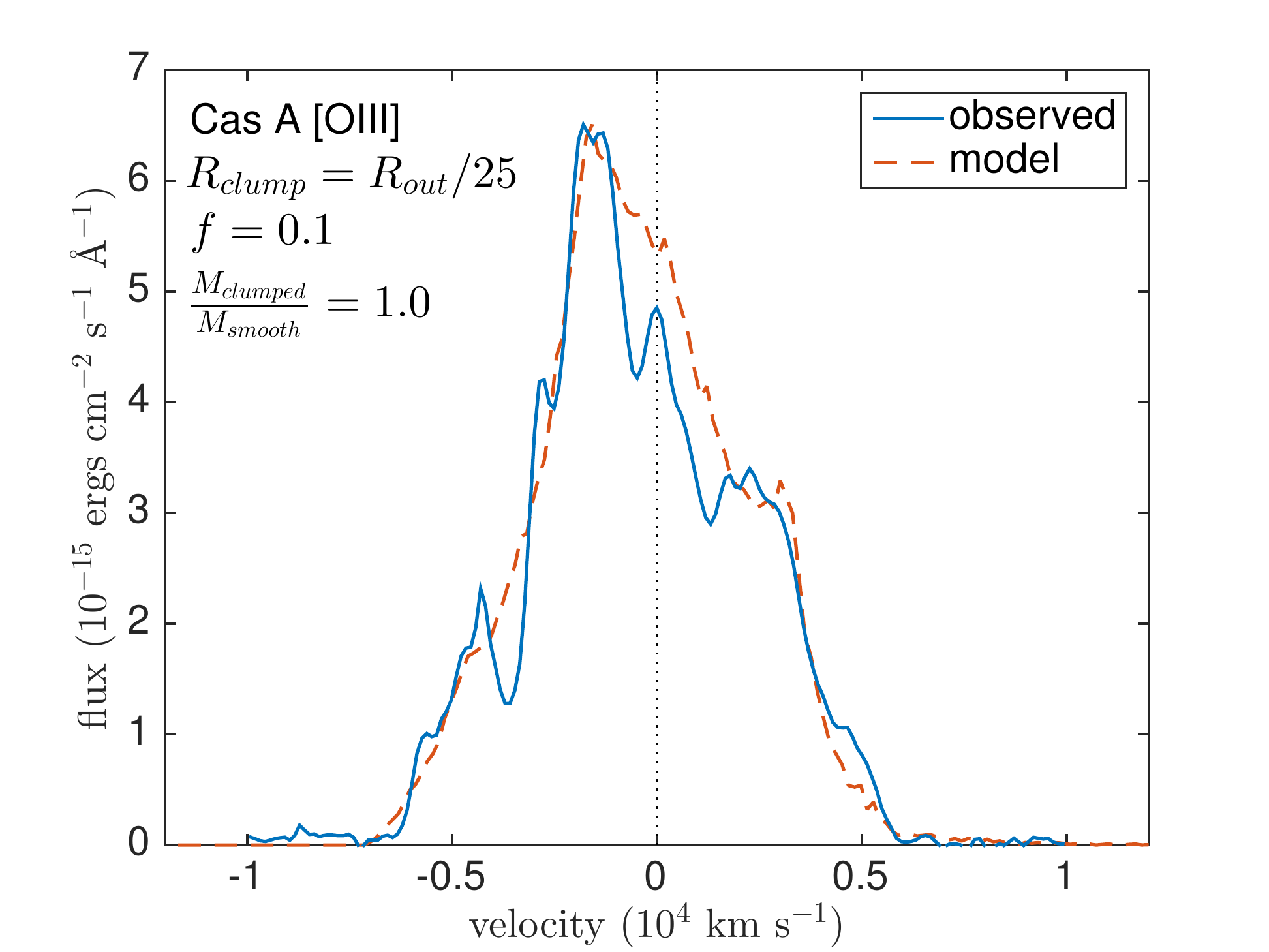}

\vspace{6mm} 
\includegraphics[scale=0.43,clip=true, trim=30 0 50 20]{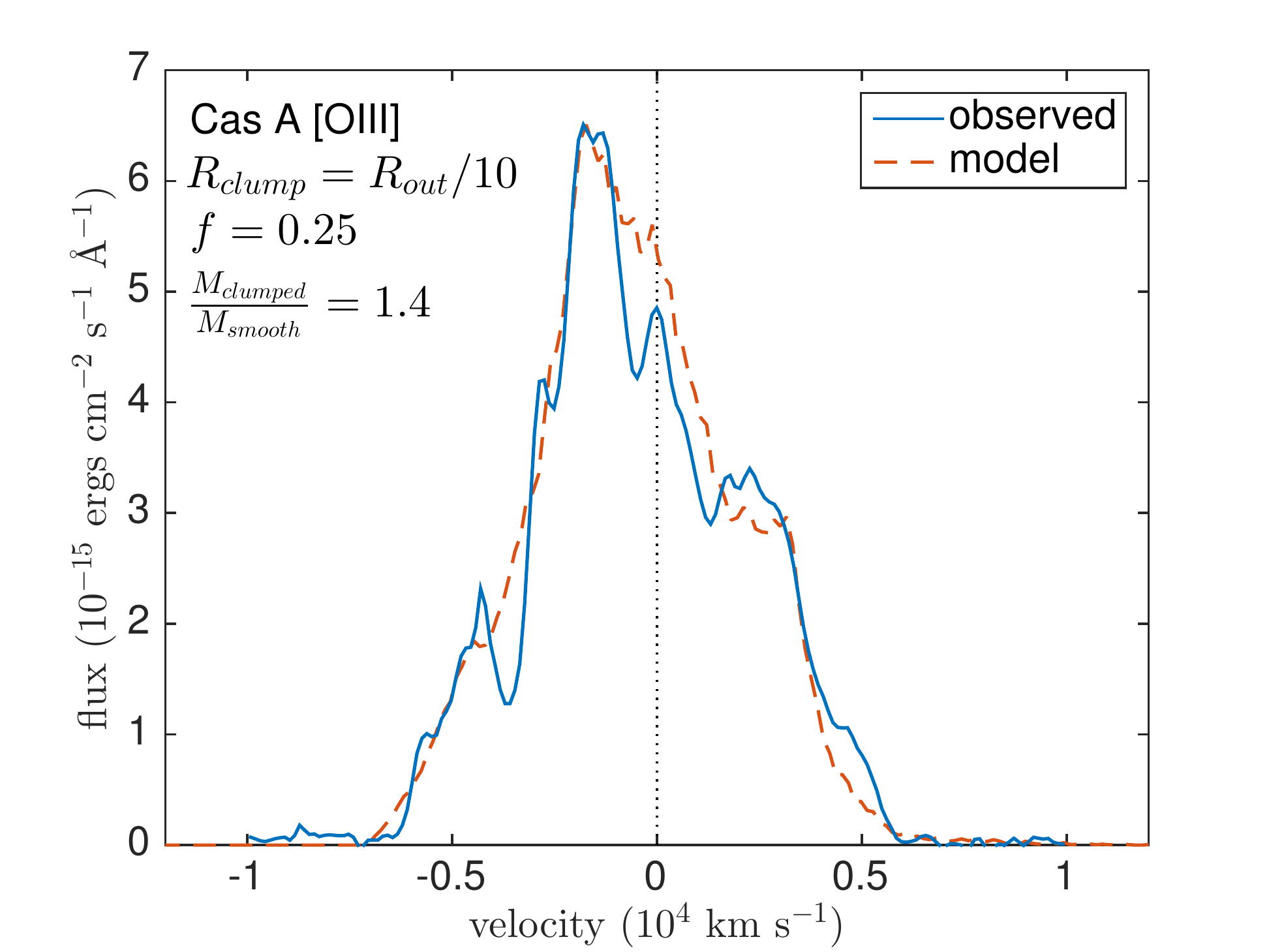} 
\includegraphics[scale=0.43,clip=true, trim=30 0 40 20]{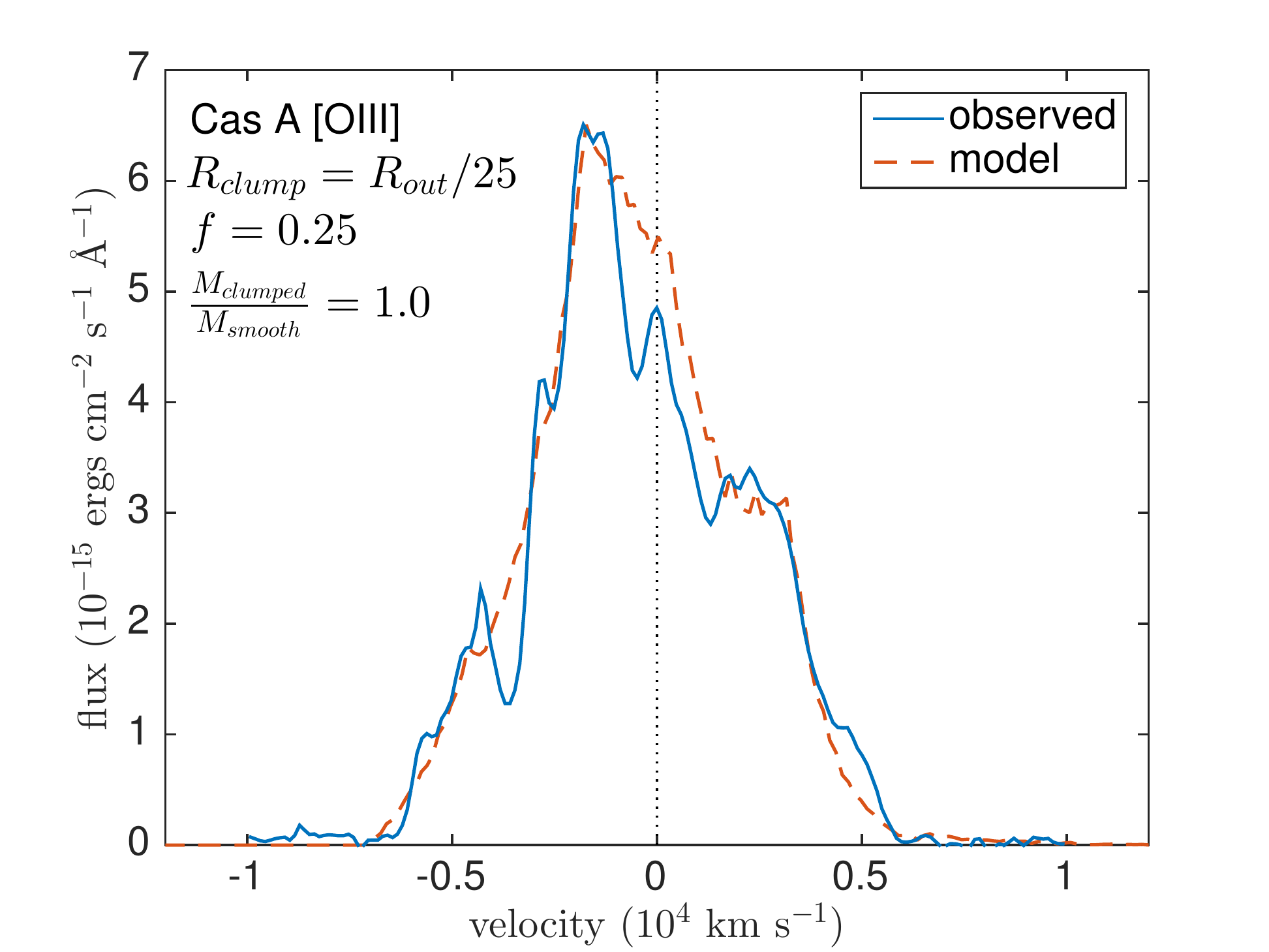}

\caption{Best clumped dust fits to the Cas~A [O~{\sc iii}]
$\lambda\lambda$4959,5007 doublet for the parameters described 
in Tables \ref{CasA_smooth_params} and \ref{CasA_clumped_dust_masses}.  
In the left column are fits to the profile using clumps with 
$R_{clump}=R_{out}/10$ and in the right column are fits using clumps with 
$R_{clump}=R_{out}/25$.  Each row uses a model that adopts a different 
clump volume filling factor with $f=0.05$ on the top, $f=0.1$ in the 
middle and $f=0.25$ on the bottom.  The model profile has been shifted 
uniformly towards the red by $+700$~km~s$^{-1}$.} 
\label{CasA_OIII_clumped} 
\end{figure*}

The factor by which the required dust mass changes relative to the 
smooth dust mass is listed in Table \ref{CasA_clumped_dust_masses} (i.e.
relative to $M_{dust}=1.1$~M$_{\odot}$ for a medium of 50\% astronomical 
silicates and 
50\% amorphous carbon -- see Table \ref{CasA_dust_masses} for other dust 
masses with different dust compositions).  Whilst clumping serves to 
increase the required dust mass in most cases, in the most extreme case it 
is still only by a factor of $\sim$3.5.  The fits for all of these cases 
are presented in Figure \ref{CasA_OIII_clumped}.

\subsection{Cas~A Discussion}

The models of Cas~A adopt a maximum expansion velocity of 
$\sim5000$~km~s$^{-1}$ which gives an outer radius of 
$5.2\times10^{18}$~cm (1.69~pc), i.e. just beyond the reverse shock radius 
of 1.57~pc \citep{Gotthelf2001}. The need to shift the profiles by either
-700~km~s$^{-1}$ or -1000~km~s$^{-1}$ in order to fit them is consistent 
with the infrared line expansion velocity asymmetry observed by 
\citet{DeLaney2010}; an 
offset of 1000~km~s$^{-1}$ applied to an originally symmetrical distribution 
between -5000~km~s$^{-1}$ and +5000~km~s$^{-1}$ results in the velocity 
range that they deduced.  \citet{DeLaney2010} derived an average 
velocity offset away from the observer of +859~km~s$^{-1}$, midway between 
the 700~km~s$^{-1}$ and 1000~km~s$^{-1}$ velocity offsets that we adopt 
for the [O~{\sc iii}] line and for the [O~{\sc i}] and [O~{\sc ii}] lines 
respectively. Additionally, from their analysis of the projected 
velocities of the optical emission line knots in Cas~A, 
\citet{Milisavljevic2013} deduced an overall asymmetry of 
+760~$\pm$~100~km~s$^{-1}$.

\begin{table} 
\caption{The ratio of clumped dust mass to  
smooth dust mass for the model whose parameters are given in Row 2 of 
Table \ref{CasA_smooth_params}, 
for clumped models with different clump widths and different clump volume 
filling factors.  The other parameters in the models were fixed at the 
values given in Table \ref{CasA_smooth_params}.} 
\centering 
\begin{tabular}{c c c c} 
\hline 
clump radius & $f=0.05$ &$f=0.1$&$f=0.25$\\ 
\hline 
$R_{out}/10$ & 3.5 & 1.9 & 1.4 \\ 
$R_{out}/25$ & 1.6 & 1.0 & 1.0 \\ 
\hline 
\end{tabular} 
\label{CasA_clumped_dust_masses} 
\end{table}

The structure of the Cas~A remnant is much more complex than 
the simple shell geometry adopted here; it
exhibits large-scale coherent structure \citep{DeLaney2010}.  
The majority of the optical ejecta are arranged in several well-defined and nearly circular filaments with diameters between approximately 30''~(0.5~pc) and 2'~(2~pc, see e.g. Figure 7 of \citet{Milisavljevic2013}).
These filaments are the likely cause of the 
noticeable bumpy substructure of the observed emission lines that we model 
here.  The models that we have presented above represent a first-order 
approximation to the geometry of Cas~A; future work will hopefully 
include a more realistic density distribution based on
line emissivity distributions derived from observations.

It is not just the geometrical structure of the Cas~A remnant that is 
complex \citep{Rho2008,Arendt2014,Biscaro2014}.   
\citet{Arendt2014} concluded that the entire spectrum of Cas~A can be 
fitted using only four dust species: Mg$_{0.7}$SiO$_{2.7}$, 
Mg$_{2.4}$SiO$_{4.4}$, Al$_2$O$_3$ and amorphous carbon.  Two of these 
species (Mg$_{0.7}$SiO$_{2.7}$, Mg$_{2.4}$SiO$_{4.4}$) are highly 
scattering and two (Al$_2$O$_3$ and amorphous carbon) are relatively 
absorbing.  This suggests that dust composition models with both silicates 
and amorphous carbon may be the most representative.  Whilst there is 
evidence for a number of grain species in the warm dust component, the 
cool dust component, found by \citet{Barlow2010} and the subject of a 
recent spatially resolved analysis by \citet{DeLooze2016}, constitutes the 
majority of the dust in Cas~A but is of still unknown composition.

The most likely dust mass of $\sim1.1$~M$_{\odot}$ for Cas~A given by our 
modelling is higher than the recent estimates of the dust mass present in 
Cas~A that were discussed at the start of this section.  However, it is 
consistent with the recent estimate \citep{DeLooze2016} of a mass of 
0.4~--~0.7~M$_{\odot}$ of cold dust in Cas~A, based on a spatially resolved 
analysis of {\em Spitzer} and {\em Herschel} imaging photometry from 
20~--~500~$\mu$m that accounted for contributions from interstellar dust and 
from supernova synchrotron and dust emission components. Further line 
profile models of Cas~A that adopt a more realistic and complex geometry 
for both the gas and dust may help to constrain its total dust mass 
further.

\section{Conclusions}

Of a sample of ten CCSNe that were still visible spectroscopically at late 
times, at least 50\% exhibited blue-shifted line profiles 
\citep{Milisavljevic2012}.  This aspect of the optical spectra of CCSNe at 
late times is most simply explained by the presence of dust in the ejecta. 
We have modelled oxygen and hydrogen line profiles in the optical spectra 
of three SNRs with ages of $\sim$16, 30 and 330 years and have found that 
we can reproduce the observed line profiles fairly well even with 
relatively simple models.  Further modelling that allows for more complex 
emission geometries may allow even better fits to be obtained. Regardless, 
it seems clear that the presence of newly-formed dust in the ejecta of 
these objects can account for the frequently seen blue-shifting of their 
line profiles.

We find that for SN~1980K a high albedo is required in order to fit the red wing of the line profiles and that  the dust composition is therefore likely dominated by silicate grains.  We obtain a dust mass in the range 0.1~--~0.3~M$_{\sun}$ at 30 years post-explosion.  In the case of SN~1993J, a dust mass of between 0.08~--~0.15~M$_{\sun}$ at 16 years post-outburst is derived based on a silicate dust model. In both cases, a clumped dust model requires approximately 1.5 times as much dust as a smooth dust model.  In the case of Cas~A, asymmetries in the ejecta require the modelled profiles to be shifted by $\sim$700~--~1000~km~s$^{-1}$.  The dust mass in Cas~A is not tightly constrained since the dust composition cannot be determined.  However, a likely composition comprises a mixture of reflective and absorbing grains.  A mixture of 50\% silicate grains and 50\% amorphous carbon grains by number yields a dust mass of 1.1~M$_{\sun}$.


Our aim throughout the modelling of these three objects has been to 
investigate the feasibility that dust causes the red-blue asymmetries 
observed in the optical line profiles from CCSNRs and then to determine 
the dust masses that cause these characteristic dust-affected line profiles.  
Whilst the derived dust masses are dependent on clumping structures 
and dust composition, at these late stages we find that significant 
dust optical depths (typically 0.5 - 2) and large dust masses
($0.1-1.1$~M$_{\odot}$) are required to account for the degree 
of blue-shifting observed.

\citet{Gall2014} brought together a number of CCSN dust mass estimates 
from the literature, based largely on infrared SED fitting, across a range of epochs up to $\sim$25 years. When they plotted dust masses as a 
function of time since the CCSN explosion, they found a steady increase. 
Separately, \cite{Wesson2015} plotted infrared- and submillimetre-based 
dust mass estimates for SN~1987A from day 615 to year 25 and found a 
similar trend of increasing dust mass with time, from 
0.001$\pm$0002~M$_\odot$ at day 615 to 0.7$\pm$0.1~M$_\odot$ at years 
23-25. The dust masses derived  by \citet{Bevan2016} from SN~1987A's line 
asymmetries up to year 10 filled in some of the gaps left by the IR/submm
SED studies and confirmed the trend of increasing dust mass with time. 
Our dust mass estimates for SN~1993J at year 16 (0.08 - 
0.15~M$_\odot$), SN~1980K at year 30 (0.12 - 0.30~M$_\odot$) and Cas~A at 
year 330 ($\sim1.1$~M$_\odot$) are broadly consistent with these trends. 
These dust masses suggest that dust formation in Type~IIb supernova 
ejecta such as SN~1993J and Cas~A is just as effective as in Type~II SN 
ejecta such as those of SN~1987A and SN~1980K. Our dust mass estimate for 
Cas~A is in broad agreement with the dust mass of 0.4~--~0.7~M$_{\odot}$ 
recently derived from a spatially resolved analysis of {\em Spitzer} and 
{\em Herschel} infrared and submillimetre imaging data for Cas~A 
\citep{DeLooze2016}.  Reasonably good multi-epoch data exist for a number of other SNe and SNRs including SN1993J and SN1979C which will provide an excellent opportunity to probe further the evolution of dust formation in these objects.

\section*{Acknowledgments}

AB's work was supported by a UK STFC Research Studentship (ST/K502406/1). MJB 
acknowledges support from STFC grant ST/M001334/1 and, since June 1st 
2016, from European Research Council (ERC) Advanced Grant SNDUST 694520.

\bibliography{80k_93j_casa}{} 
\bibliographystyle{mnras}

\end{document}